\documentclass[%
 reprint,
 amsmath,amssymb,
 aps,
]{revtex4-2}

\usepackage{graphicx}
\usepackage{dcolumn}
\usepackage{bm}

\usepackage{amsmath,amssymb,amsfonts,amsthm,bm,bbm,cancel,wasysym}
\usepackage{epsfig,graphics,graphicx,epstopdf,caption,subcaption}
\graphicspath{{Charts/}}
\usepackage{array,booktabs,colortbl,colordvi,multirow}
\usepackage{colordvi,color,xcolor}
\usepackage{hyperref}
\usepackage{rotating}
\usepackage{comment}
\usepackage{feynmp-auto}
\usepackage{verbatim}
\usepackage{tikz}
\usetikzlibrary{calc, shapes, arrows, decorations.pathmorphing, decorations.pathreplacing, 3d, shapes.multipart}

\tikzstyle{particle} = [diamond, draw, text centered, rounded corners, minimum height = 2em]
\tikzstyle{coupling} = [circle, draw, text centered, rounded corners, minimum height = 2em]
\tikzstyle{action} = [rectangle, draw, text centered, rounded corners, minimum height = 2em]
\tikzstyle{input} = [circle, draw, text centered, minimum height = 2em]
\tikzstyle{graph} = [rectangle, draw, rounded corners, text centered, minimum height = 2em]
\tikzstyle{mlp} = [rectangle, draw, text centered, minimum height = 2em]
\tikzstyle{pooling} = [diamond, draw, text centered, minimum height = 2em]
\usepackage{setspace}
\usepackage{url}
\usepackage[percent]{overpic}
\usepackage{slashed}
\usepackage{xspace}
\usepackage{fullpage}
\newskip\zatskip \zatskip=0pt plus0pt minus0pt
\def\matth{\mathsurround=0pt}

\def\atversim#1#2{\lower0.7ex\vbox{\baselineskip\zatskip\lineskip\zatskip
  \lineskiplimit 0pt\ialign{$\matth#1\hfil##\hfil$\crcr#2\crcr\sim\crcr}}}

\newif\ifdiagrams
\diagramstrue
\ifdiagrams

\else
  \excludecomment{fmffile}
\fi

\usepackage{hyperref} 
\hypersetup{
    colorlinks=true,       
    linkcolor=red,          
    citecolor=blue,        
    filecolor=magenta,      
    urlcolor=blue           
}
\usepackage[all]{hypcap} 

\begin{document}

\preprint{APS/123-QED}

\title{Towards Beyond Standard Model Model-Building with Reinforcement Learning on Graphs}

\author{George N. Wojcik}
\email{gwojcik@wisc.edu}
\author{Shu Tian Eu}
\email{eu@wisc.edu}
\author{Lisa L.~Everett}%
\email{leverett@wisc.edu}
\affiliation{%
Department of Physics, University of Wisconsin-Madison, Madison, WI 53706, USA\\
}%

\date{\today}

\begin{abstract}
We provide a framework for exploring physics beyond the Standard Model with reinforcement learning using graph representations of new physics theories. The graph structure allows for model-building without {\it a priori} specifying definite numbers of new particles. As a case study, we apply our method to a simple class of theories involving vectorlike leptons and a dark $U(1)$ inspired by the portal matter paradigm. Using modern policy gradient methods, the agent successfully explores a model space consisting of both continuous and discrete parameters and identifies consistent theories. The minimal models found include both known and previously unstudied examples that can accommodate the muon anomalous magnetic moment and satisfy precision electroweak and flavor constraints. The method represents a step forward in enabling an automated model-building process for physics beyond the Standard Model.   
\end{abstract}

\maketitle


\section{\label{sec:intro}Introduction}

In the current era in which models of physics beyond the Standard Model (BSM) have been experimentally constrained at unprecedented levels, identifying means for more comprehensive explorations of the space of potentially viable BSM theories, which encompasses all theories which contain the Standard Model effective field theory and possess only additions which avoid present experimental constraints, is of increasing importance. Reinforcement learning (RL) provides one such method for an automated exploration of such large spaces for which viable points are sparsely distributed. In an RL scan an agent is trained to recommend actions to modify a model in ways that maximize its expected reward, which here is by design correlated to some metric of the theoretical or phenomenological viability of the model. 

RL has already demonstrated promise in the field of BSM model building for frameworks that have a finite number of discrete parameters, for example in probing large spaces in string theory \cite{Halverson:2019tkf}, and identifying viable Froggatt-Nielsen charges \cite{Harvey:2021oue,Nishimura:2020nre}. However, if reinforcement learning scans are to be generalized effectively to a broader class of BSM model building problems, it is crucial that the procedure be adapted to scanning over spaces of models where the BSM particle content, and therefore the feature dimensionality of the subspace, is variable.

To this end, we present a procedure in which BSM theories are represented as graphs and use graph neural networks. Our procedure can be used to explore any class of four-dimensional BSM theories with a finite but otherwise arbitrary and not pre-specified number of fields, essentially only excluding theories with infinite towers of states such as Kaluza-Klein theories. The theorist would then just need to specify the symmetry group and the possible group representations of the particles in the theory. As a proof-of-concept, we apply our method to a subclass of BSM models with vector-like leptons and a dark $U(1)$ gauge group, inspired by the portal matter framework \cite{Rizzo:2018vlb,Rizzo:2022lpm}. We impose a minimal set of {\it a priori} assumptions on the model space, including making no prior stipulations enforcing flavor conservation. Upon imposing experimental constraints, including the muon anomalous magnetic moment discrepancy \cite{Muong-2:2006rrc,Muong-2:2021ojo,Muong-2:2023cdq,Aoyama:2020ynm,Aoyama:2012wk,Aoyama:2019ryr,Czarnecki:2002nt,Gnendiger:2013pva,Davier:2017zfy,Keshavarzi:2018mgv,Colangelo:2018mtw,Hoferichter:2019mqg,Davier:2019can,Keshavarzi:2019abf,Kurz:2014wya,Melnikov:2003xd,Masjuan:2017tvw,Colangelo:2017fiz,Hoferichter:2018kwz,Gerardin:2019vio,Bijnens:2019ghy,Colangelo:2019uex,Blum:2019ugy,Colangelo:2014qya}, and flavor and electroweak precision constraints, and incentivizing simplicity, our procedure yields not only minimal constructions that have already been explored \cite{Wojcik:2022woa}, but also alternatives that have not to our knowledge been previously studied in the literature.  

This paper is structured as follows. In Section II, we describe the representation of BSM theories as graphs and discuss graph neural networks. We turn in Section III to a discussion of our reinforcement learning environment. In Section IV, we present the results of the RL scan for vector-like lepton theories considered. The discussion and conclusions are provided in Section V. This paper summarizes our main results; the reader is referred to our companion long paper \cite{Wojcik:2024xxx} for a comprehensive discussion.

\section{\label{sec:model-setup} BSM Graph Representations}

Our procedure is based on leveraging the utility of using graphs in representing arbitrary BSM theories, with learning tasks accomplished through a graph neural network. Mathematically, a graph consists of nodes and the edges which connect them, where each node is described by a feature vector, and each edge consists of the pair of connected nodes and a corresponding feature vector for the edge that characterizes the coupling.  In our BSM graph grammar, each node in a graph will represent either a field or an interaction term. Edges connect field nodes to coupling terms, allowing interaction terms with arbitrary numbers of different fields to be represented with edges that connect only pairs of particles. 

Here we are interested in four-dimensional BSM theories with an arbitrary but finite number of different fields, which have representations under Lorentz symmetry and any proposed internal symmetries. With the assumption that there is a finite number of {\it distinct} group representations appearing in the model, and that there is a finite maximum mass dimension for the allowed interaction terms that are written down, it can be shown that the full set of nodes and edges, including the elements of their feature vectors, is finite \cite{Wojcik:2024xxx}.

For the class of models we consider here, the new states are electroweak doublet or singlet vector-like leptons with $U(1)_D$ charges of $\pm 1$ or 0. We denote these states by $L_q$ and $E_q$, respectively, where $q$ is its $U(1)_D$ charge, with mass parameters $M^{L(E),0 (\pm)}$ (we choose to work in a basis where there are no mixed mass parameters between the SM leptons and the new vector-like leptons). These states interact with the SM leptons via Yukawa couplings to the SM Higgs $h$ or to a dark Higgs field $h_D$, which breaks the $U(1)_D$. Since the $U(1)_D$ breaking is at the sub-GeV scale in portal matter models, we will not need to include the dark Higgs or the dark photon explicitly as nodes in our BSM graphs, since for observable quantities of interest for the vector-like leptons these sub-GeV mass scales do not directly appear.

The vector-like leptons couple to the SM leptons through the Yukawa couplings $\lambda^\pm_{L (E)}$, which couple $L_\pm$ ($E_\pm$) to the SM leptons via $h_D$, and $y^0_{L(E)}$, which couple $L^0$ ($E^0$) to the SM leptons via $h$. The vector-like leptons can also have Yukawa interactions with each other in various ways: $y^\alpha_{LE,EL}$ denotes the Yukawa couplings of $L_q$ and $E_q$ with the SM Higgs ($\alpha=\{0,\pm \}$), and $\lambda^{0\pm,\pm 0}_{L(E)}$ are the Yukawa couplings of vector-like doublets (singlets) of dark charge $\pm 1$ to those with dark charge 0.  
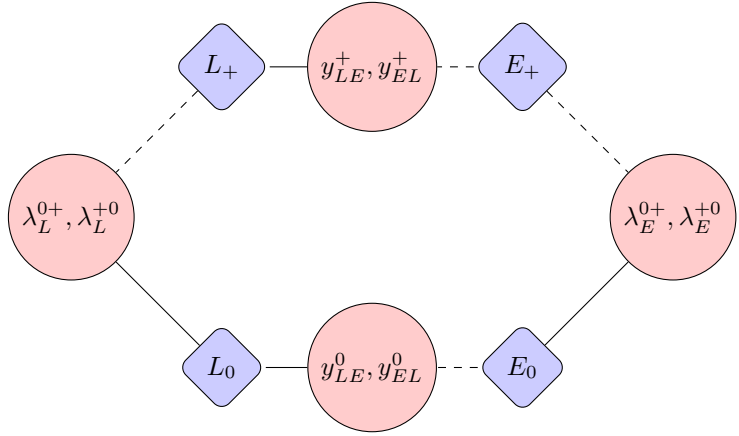
\begin{figure*}[!ht]
    \begin{tikzpicture}[scale=1]
        \node at (0,4) [particle, fill=blue!20] (lp-id) {$L_+$};
        \node at (0,0) [particle, fill=blue!20] (l0-id) {$L_0$};
        \node at (4,4) [particle, fill=blue!20] (ep-id) {$E_+$};
        \node at (4,0) [particle, fill=blue!20] (e0-id) {$E_0$};
        \node at (-2,2) [coupling, fill=red!20] (lamLP-id) {$\lambda^{0+}_L, \lambda^{+0}_L$};
        \node at (2,4) [coupling, fill=red!20] (yLEP-id) {$y^+_{LE},y^+_{EL}$};
        \node at (2,0) [coupling, fill=red!20] (yLE0-id) {$y^0_{LE},y^0_{EL}$};
        \node at (6,2) [coupling, fill=red!20] (lamEP-id) {$\lambda^{0+}_E, \lambda^{+0}_E$};
        \draw[-] (l0-id) -- (lamLP-id);
        \draw[dashed] (lp-id) -- (lamLP-id);
        \draw[-] (lp-id) -- (yLEP-id);
        \draw[dashed] (ep-id) -- (yLEP-id);
        \draw[-] (l0-id) -- (yLE0-id);
        \draw[dashed] (e0-id) -- (yLE0-id);
        \draw[-] (e0-id) -- (lamEP-id);
        \draw[dashed] (ep-id) -- (lamEP-id);
    \end{tikzpicture}
    \caption{A visual depiction of a graph in the class of theories explored with four total BSM fermions (an isospin singlet and a doublet that are uncharged under $U(1)_D$, and an isospin singlet and a doublet that have a dark charge of $+1$), following the graph grammar of Table \ref{tab:rep-A}. Diamonds denote particle nodes, which contain the particle mass and Yukawa couplings to SM particles as features, and circles denote Yukawa couplings. A solid line denotes an edge of type $e_1$, while a dashed line denotes an edge type of $e_2$.}
    \label{fig:vll-graph-A}
\end{figure*}

Our graph grammar for such theories is shown in Table~\ref{tab:rep-A}. As seen, the graph grammar represents theories as a heterogeneous graph with two node types of different feature dimensionality:  particles and couplings. Discrete node features are used to identify the electroweak representation and dark $U(1)$ charge of particle nodes, and a discrete node feature is used to distinguish between dark Higgs and SM Higgs Yukawa couplings among the coupling nodes\footnote{It is in principle possible to express an identical graph grammar using a traditional homogeneous graph, without differing node types. For more details on the subtleties of this construction we refer the reader to \cite{Wojcik:2024xxx}.}.  Note that the Yukawa coupling terms between two vector-like fields must treat the two incoming fields differently, otherwise terms such as the $y_{EL}$ couplings would not be distinguished from the $y_{LE}$ couplings. Therefore, for each different type of interaction node we consider, we must have two different edge types, as shown. The relations between various Yukawa couplings among different particles is shown graphically in Figure~\ref{fig:vll-graph-A}.

\begin{table*}[t]
    \centering
    \begin{tabular}{| c | c | c | c | c |}
    \hline
    \multicolumn{3}{| c |}{Nodes} & \multicolumn{2}{| c |}{Edges}\\
    \hline
    Node Type & Field/Coupling & Feature Vector & Edge Type & Particle $F$ to Coupling $\{ g_1, g_2 \}$\\
    \hline
    \multirow{4}{*}{Particle} & $L_0$ & $\{ M^{L,0}, \vec{y}^0_L, 0, 0 \}$ & \multirow{4}{*}{$e_1$} & \multirow{4}{*}{$g_1 \overline{F} P_R \square + g_2 \overline{\square} P_R F$}\\
    & $L_\pm$ & $\{ M^{L, \pm}, \vec{\lambda}^\pm_L, \pm 1, 0 \}$ & &\\
    & $E_0$ & $\{ M^{E, 0}, \vec{y}^0_E, 0, 1 \}$ & &\\
    & $E_\pm$ & $\{ M^{E, \pm}, \vec{\lambda}^\pm_E, \pm 1, 1 \}$ & &\\
    \hline
    \multirow{4}{*}{Coupling} & $y^{0}_{LE}, \, y^{0}_{EL}$ & $\{ y^{0}_{LE}, y^{0}_{EL}, 0 \}$ & \multirow{4}{*}{$e_2$} & \multirow{4}{*}{$g_1 \overline{\square} P_R F + g_2 \overline{F} P_R \square$}\\
    & $y^{\pm}_{LE}, \, y^{\pm}_{EL}$ & $\{ y^{\pm}_{LE}, y^{\pm}_{EL}, 0 \}$ & &\\
    & $\lambda^{0 \pm}_L, \, \lambda^{\pm 0}_L$ & $\{ \lambda^{0 \pm}_L, \lambda^{\pm 0}_L, 1 \}$ & &\\
    & $\lambda^{0 \pm}_E, \, \lambda^{\pm 0}_E$ & $\{ \lambda^{0 \pm}_E, \lambda^{\pm 0}_E, 1 \}$ & &\\
    \hline
    \end{tabular}
    \caption{The graph grammar used to represent models of the class explored. $\vec{y}$ and $\vec{\lambda}$ are three-component vectors which describe the Yukawa couplings between vector-like fermions and the SM fields.}\label{tab:rep-A}
\end{table*}

\section{Reinforcement Learning Environment}\label{sec:environment}

\begin{figure*}
    \centering
    \begin{tikzpicture}
        \node at (5,4) [action, fill=blue!20] (m-id) {Master Action};
        \node at (0,2) [action, fill=blue!20] (mc-id) {Select Node};
        \node at (0,1) [action, fill=blue!20] (c-id) {Select Parameter};
        \node at (0,0) [action, fill=red!20] (p-id) {Choose Modification};
        \node at (5,2) [action, fill=blue!20] (ad-id) {Select Representation};
        \node at (5,1) [action, fill=red!20] (ac-id) {Choose Internal Parameters};
        \node at (5,0) [action] (y-id) {Randomly Generate Coupling Nodes};
        \node at (10,2) [action, fill=blue!20] (dc-id) {Select Particle};
        \draw[->] (m-id) -- (mc-id) node[midway,above left] {Modify Node};
        \draw[->] (mc-id) -- (c-id);
        \draw[->] (c-id) -- (p-id);
        \draw[->] (m-id) -- (ad-id) node[midway] {Add Particle};
        \draw[->] (ad-id) -- (ac-id);
        \draw[->] (ac-id) -- (y-id);
        \draw[->] (m-id) -- (dc-id) node[midway,above right] {Delete Particle};
    \end{tikzpicture}
    \caption{A visual summary of the structure of the action space in our reinforcement learning environment. Each blue box represents a discrete action, while red particles denote continuously-parameterized actions. The sole white box denotes random sampling with a uniform prior (thus not informed by a learned policy).}
    \label{fig:action}
\end{figure*}
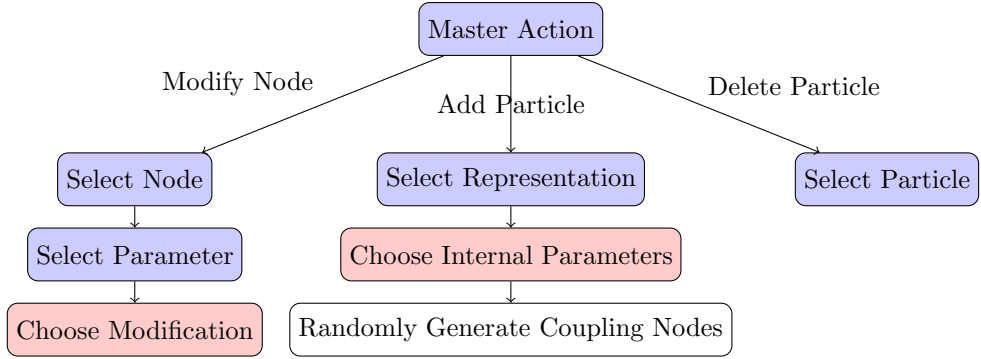

We use the graph grammar we have constructed to form the core of a reinforcement learning environment through which we will explore the space of these vector-like lepton theories. The problem of reinforcement learning is modelled as a Markov decision process, consisting of a state space $\mathcal{S}$, a space of actions $\mathcal{A}$, and a policy $\pi_\theta$. Given a state $s \in \mathcal{S}$, the policy $\pi_\theta$ will assign probabilities to different actions $a \in \mathcal{A}$ which will transform the state into a new one. The parameters $\theta$, in our case the weights of a neural network, define the behavior of this policy. Desirable states for the model are given positive numerical values known as rewards; following a training algorithm the parameters $\theta$ are optimized to maximize the accumulated reward that the agent will achieve over some trajectory of actions. We define these trajectories as episodes, which will comprise the set of states produced by a series of modifications of an original randomly-sampled model, until either a model which we have defined as ``terminal'' (namely, phenomenologically promising) is produced or the agent has taken 250 actions.

In our context, the state space consists principally of a graph representing a BSM model, in addition to graph metadata counting the number of vector-like leptons of each species that are present in the model. The agent will be rewarded based on producing models which have a larger log-likelihood than the SM, based on the precision observables of Table \ref{tab:observables}, and for achieving this likelihood with a minimal number of BSM particles. We find good results for a model evaluation metric inspired by the Akaike Information Criterion (AIC), given as
\begin{align}\label{eq:model-eval-metric}
    K(s) \equiv \log \bigg( \frac{L(\textrm{Data} | s)}{L(\textrm{Data} | \textrm{SM}) } \bigg) - k \times n_{\textrm{particles}},
\end{align}
where $L(\textrm{Data} | s)$ is the likelihood of the model state $s$ given the precision observables of Table \ref{tab:observables}, $L(\textrm{Data} | \textrm{SM})$ is the same for the SM, $n_{\textrm{particles}}$ is the total number of vector-like leptons which are present in the model, and $k$ is a constant, which for our trials we take to be $k = 0.5$. At each step, the agent receives rewards based on improving on the highest score that it has received so far in the episode. These rewards take the form
\begin{align}
    &R (K_t, K_{\textrm{max}}) = \\
        &\theta(K_t - K_{\textrm{max}}) \sum_{j} [j \theta(K_t - m_i) - j \theta(K_{\textrm{max}}-m_i)], \nonumber\\
    &\vec{m} \equiv [-10, -5, -2, 1, 2, 3, 4, 5, 6, 7, 8]. \nonumber
\end{align}
To ensure that our expected rewards follow the Markov property, the state space is supplemented with $K_{\textrm{max}}$ at each step. Similarly, because the number of steps in an episode is truncated, we include the number of steps that the agent has already taken as an additional state variable. In addition to the intermediate rewards, the agent receives a reward of $+100$ for achieving a terminal state which is phenomenologically promising enough to conclude the episode early. We define a terminal state as
\begin{align}
\label{eq:terminal-score}
    &K(s_{\textrm{terminal}}) \geq 9 - 2 k,\\
    &k = 0.5 \rightarrow K(s_{\textrm{terminal}}) \geq 8.\nonumber
\end{align}

\begin{table*}[]
    \centering
    \begin{tabular}{| c | c | c | c |}
    \hline
    Observable & Definition & Value & Source \\
    \hline
    $M_W$ & $W$ boson mass & $80.377 \pm 0.012 \; \textrm{GeV}$ & \cite{Workman:2022ynf,LHCb:2021bjt,CDF:2022hxs,ATLAS:2017rzl,CDF:2012gpf,D0:2012kms,DELPHI:2008avl,OPAL:2005rdt,L3:2005fft,ALEPH:2006cdc,D0:2002fhu,CDF:2000gwd}\\
    \hline
    $BR(W \rightarrow e \nu)$ & \multirow{3}{*}{$W$ partial width} & $0.1071 \pm 0.0016$ & \multirow{3}{*}{\cite{Workman:2022ynf,CMS:2022mhs,OPAL:2007ytu,DELPHI:2003ftu,L3:2004lwm,ALEPH:2004dmh}}\\
    $BR(W \rightarrow \mu \nu)$ & & $0.1063 \pm 0.0015$ & \\
    $BR(W \rightarrow \tau \nu)$ & & $0.1138 \pm 0.0021$ & \\
    \hline
    $R_e$ & \multirow{3}{*}{$Z$ partial width ratio (hadrons to leptons)} & $20.804 \pm 0.050$ & \multirow{3}{*}{\cite{Workman:2022ynf,OPAL:2000ufp,DELPHI:2000wje,L3:2000vgx,ALEPH:1999smx}}\\
    $R_\mu$ & & $20.784 \pm 0.034$ & \\
    $R_\tau$ & & $20.764 \pm 0.045$ & \\
    \hline
    $A_e$ & \multirow{3}{*}{$Z$ pole electron asymmetry parameter} & $0.1515 \pm 0.0019$ & \cite{Workman:2022ynf,OPAL:2001brm,SLD:2000ujp,ALEPH:2001uca,DELPHI:1999yne,L3:1998oan,SLD:1996gjt,SLD:1994kvj}\\
    $A_\mu$ & & $0.142 \pm 0.015$ & \cite{Workman:2022ynf,SLD:2000ujp}\\
    $A_\tau$ & & $0.143 \pm 0.004$ & \cite{Workman:2022ynf,OPAL:2001brm,SLD:2000ujp,ALEPH:2001uca,DELPHI:1999yne,L3:1998oan}\\
    \hline
    $A_{FB}^{(0,e)}$ & \multirow{3}{*}{$Z$ pole forward-backward asymmetry} & $0.0145 \pm 0.0025$ & \multirow{3}{*}{\cite{Workman:2022ynf,OPAL:2000ufp,DELPHI:2000wje,L3:2000vgx,ALEPH:1999smx}}\\
    $A_{FB}^{(0,\mu)}$ & & $0.0169 \pm 0.0013$ & \\
    $A_{FB}^{0,\tau}$ & & $0.0188 \pm 0.0017$ & \\
    \hline
    $\Delta a_e$ & \multirow{2}{*}{lepton anomalous magnetic moment
    }& $(-8.8 \pm 3.6) \times 10^{-13}$ & \cite{2018Sci...360..191P}\\
    $\Delta a_\mu$ & & $(2.51 \pm 0.59) \times 10^{-9}$ & \cite{Muong-2:2006rrc,Muong-2:2021ojo,Aoyama:2020ynm,Aoyama:2012wk,Aoyama:2019ryr,Czarnecki:2002nt,Gnendiger:2013pva,Davier:2017zfy,Keshavarzi:2018mgv,Colangelo:2018mtw,Hoferichter:2019mqg,Davier:2019can,Keshavarzi:2019abf,Kurz:2014wya,Melnikov:2003xd,Masjuan:2017tvw,Colangelo:2017fiz,Hoferichter:2018kwz,Gerardin:2019vio,Bijnens:2019ghy,Colangelo:2019uex,Blum:2019ugy,Colangelo:2014qya}\\
    \hline
    $y_\mu$ & muon Yukawa coupling & $1.12 \pm 0.2$ & \multirow{2}{*}{\cite{Workman:2022ynf,ATLAS:2022vkf,CMS:2022dwd}}\\
    $y_\tau$ & $\tau$ Yukawa coupling & $0.94 \pm 0.07$ & \\
    \hline
    $BR(\mu \rightarrow e \gamma)$ & $\mu \rightarrow e \gamma$ branching fraction & $< 4.2 \times 10^{-13}$ & \cite{MEG:2016leq}\\
    $BR(\tau \rightarrow e \gamma)$ & $\tau \rightarrow e \gamma$ branching fraction & $< 3.3 \times 10^{-8}$ & \cite{BaBar:2009hkt}\\
    $BR(\tau \rightarrow \mu \gamma)$ & $\tau \rightarrow \mu \gamma$ branching fraction & $< 4.2 \times 10^{-8}$ & \cite{Belle:2021ysv}\\
    $\Gamma^{\textrm{conv}}_{\textrm{Au}} / \Gamma^{\textrm{capt}}_{\textrm{Au}}$  & $\mu-e$ conversion in gold nuclei & $< 7 \times 10^{-13}$ & \cite{SINDRUMII:2006dvw}\\
    \hline
    \end{tabular}
    \caption{The physical observables used in our analysis. Upper limits are quoted as 90\% CL bounds.}
    \label{tab:observables}
\end{table*}

The action space $\mathcal{A}$ consists of a series of hierarchical actions, in which top-level decisions about modifications that the agent can make are parameterized by a number of sub-decisions. The three classes of actions are as follows:

\begin{itemize}
    \item \textbf{Delete a particle:} If selected, the agent has one associated sub-action: Specifying which in the model to delete. Once the particle is deleted, all coupling nodes that are associated with it are automatically deleted as well.
    \item \textbf{Add a particle:} If selected, the agent must then make discrete decisions about the electroweak representation (singlet or doublet) of the new particle, as well as its charge under the dark $U(1)_D$ group ($-1$, $0$, or $+1$). Then, the agent must select the four continuous features describing a particle node (its mass and its Yukawa couplings to the three charged SM leptons). Yukawa couplings between the new particle and BSM particles already in the model are sampled randomly from $O(1)$ parameters, and may be modified by agent actions.
    \item \textbf{Modify a continuous feature on an existing node:} If selected, the agent will first select a node (either coupling or particle) to modify, and then the parameter of that node that it will modify. Finally, it selects a numerical modification to be made to that parameter.
\end{itemize}

The structure of the action space of these three classes of actions described is depicted in Figure~\ref{fig:action}. Clearly, the action space is highly hierarchical, a structure which is not supported by all reinforcement learning algorithms. However, the H-PPO reinforcement learning algorithm \cite{2019arXiv190301344F} is expressly designed to learn policies over such an action space, and so we select it when implementing our environment.

Our policy $\pi_\theta$ over these actions will be parameterized by a neural network which takes the environment state as an input, and outputs both a prediction for the long-term rewards that the agent can expect from the state over the course of the episode, and probability distributions from which we can sample actions. For each continuous action parameter, the agent must output both a mean and a variance, which together will define a normal distribution from which a continuous action parameter will be sampled. For each discrete action parameter, the agent must output a log-probability associated with each choice, which will define a probability over the discrete options from which the action will be sampled. In the terminology of graph neural networks, different action parameters are node-level and graph-level outputs, depending on the action parameter in question. In order to accomodate large hierarchies between parameters while permitting the neural network to process only $O(1)$ inputs, we also take the liberty of representing magnitudes such as the model evaluation score and the various Yukawa couplings in scientific notation to the agent's neural network, so, for example, a feature with a value $0.02$ will be passed as a length-two vector $\{ 2., -2 \}$, where the first parameter is a continuous $O(1)$ value and the second is a discrete exponent. In turn, this scientific notation representation allows the action space to include modifications of either the exponential or continuous part of model parameters. 

For details involving our neural network implementation, including the manner in which supplementary state data (such as the current maximum episode score and the number of steps the agent has already taken) are handled, we refer the reader to our longer companion paper \cite{Wojcik:2024xxx}.

\section{Results}
For our experiments, we implement a customized learning environment built with the Python library gymnasium \cite{towers_gymnasium_2023}, and implement our graph neural network using the library Pytorch Geometric \cite{2019arXiv190302428F}. To produce reinforcement learning trajectories, we randomly sample 32 parallel environments from the model space. These samples (and the environment in general) are subject to some mild restrictions: To simulate current collider limits on vector-like lepton production and limit our constructions to regions in which these leptons might be produced at realistic future colliders, we limit our model space by requiring all vector-like leptons to have a mass of $\geq 1.5 \; \textrm{TeV}$ (easily consistent with current LHC constraints on vector-like leptons \cite{OsmanAcar:2021plv,ATLAS:2015qoy,ATLAS:2023sbu,Guedes:2021oqx,Kawamura:2023zuo,Bhattiprolu:2019vdu}) and require that these particles be no heavier than $7 \; \textrm{TeV}$ (which might be achievable via pair production at a $\sqrt{s} \sim 14 \; \textrm{TeV}$ future lepton collider). For simplicity, we assume that all Yukawa coupling constants are real, but may have either positive or negative sign, and, we generally restrict them to have a magnitude between $10^{-10}$ and $10$. SM Higgs Yukawa coupling constants which mix the SM leptons with zero dark-charge vector-like leptons are required to have magnitude between $10^{-12}$ and $0.1$ to avoid parameter regimes in which the SM leptons' mixing with the BSM states can become large. Finally, although our graph architecture permits us to consider models with arbitrarily large particle content, we place a modest limit on the number of additional BSM particles because the complexity of the numerical likelihood calculation will increase polynomially with larger numbers of such particles, simply in order to slightly accelerate computation time. We also here incentivize simplicity by defining the terminal state to have minimal number of BSM states, which here will be at most two new vector-like states, consistent with Eq.~(\ref{eq:terminal-score}).

The agent then evolves the model according to its policy for 50 time steps before being trained according to the H-PPO algorithm, after which point it will continue. Episodes that have not attained a terminal state will end after 250 total steps. We allow each agent to train simultaneously over the 32 parallel environments, and continue until 1000 rounds of 50 steps each have been completed-- in total, then, the agent samples 1.6 million models over the course of its training. We record all terminal states that the agent produces over the course of its training. To mitigate the significant variance in performance due to initial trajectories (which in turn heavily depend on the initial training model states, the agent's initial weight parameters, and even the random sampling from the agent's policy probabilities early in training), we also perform each experiment 10 independent times with 10 independent initializations.

In our companion paper \cite{Wojcik:2024xxx}, we perform a scan over a number of training hyperparameters and environment architecture choices, in order to explore the factors which influence the performance of this methodology. Here, however, we shall limit ourselves to simply quoting the notable physics results which have emerged from our scan, in particular the variety of different models (that is, models with differing particle content) for which the agent has discovered terminal states. In a crucial nod to this methodology's potential viability as a model builder, however, we note here that although we have performed a large number of reinforcement learning scans over the course of our experiments, we have also discovered a number of hyperparameter configurations that are capable of fully exhausting the set of viable distinct sets of particle content over the course of a single scan, which do not appear to be overly sensitive to extremely precise parameter tuning.

 In our scans, the reinforcement learning agents have identified six distinct sets of minimal model particle content, each with two BSM particles. The six models all rely on the dark photon and dark Higgs contributions to the muon anomalous magnetic moment at one loop, which experiences a chiral enhancement due to a Yukawa coupling (either with the SM Higgs or the dark Higgs) between different vector-like leptons, in the manner discussed in \cite{Wojcik:2022woa}. These models can be classified as follows:
\begin{itemize}
\item The two new particles are vector-like leptons with the same $U(1)_D$ charge but different electroweak representations: $\{L_+,E_+\}$ (model `a') and $\{L_-,E_-\}$ (model `b').
\item The two new particles are vector-like leptons with the same electroweak representations, but different $U(1)_D$ charges, one with dark charge of $\pm 1$ and one with $0$: $\{L_0,L_+\}$ (model `c'), $\{L_0,L_-\}$ (model `d'), $\{E_0,E_+\}$ (model `e'), and $\{E_0,E_-\}$ (model `f').
\end{itemize}
 In models `a' and `b,' we observe the chiral enhancement mechanism discussed previously in \cite{Wojcik:2022woa}, in that new vector-like particles with the same dark charge but different electroweak representations can have significant Yukawa couplings to the SM Higgs.
In models `c'-`f,' the chiral enhancement mechanism arises from a different source, in which it is the Yukawa couplings of vector-like leptons with the same electroweak representations but different dark charges that play a dominant role. In for example theories of the classes `e' and `f', the correction to the SM prediction for the muon anomalous magnetic moment can be estimated to be of the form
\begin{align}
    a_{\mu}^{e,f} \approx - \frac{m_\mu^2}{16 \sqrt{2} \pi^2} \frac{M^{E,\pm}}{m_\mu} \frac{\lambda_{E, \mu}^{\pm} \lambda_E^{\pm 0} v_D^2}{(M^{E, \pm})^2} \frac{y_{E, \mu}^0 v}{M^{E, 0}}.
\end{align}
In this case, the left-handed SM leptons mix primarily with the zero dark charge vector-like leptons through couplings with the SM Higgs, while right-handed SM leptons mainly mix with the $\pm 1$ dark charge vector-like lepton. The chiral enhancement then results at leading order from the mixing between the two vector-like lepton species, from the dark Higgs Yukawa coupling $\lambda^{\pm 0}_E$. 

The different models not only can have different phenomenological implications, but also suggest different possible embeddings in higher-scale theories. As is well known, the production cross sections for vector-like leptons with different electroweak representations can vary significantly, with vector-like electroweak doublets having larger production cross sections that are more strongly constrained by collider searches than those of electroweak singlets \cite{Rizzo:2022lpm,OsmanAcar:2021plv}. In addition, the decay channels of vector-like leptons with nonzero dark charge are dominated by channels with dark photon rather than electroweak gauge boson emission, which generally have stronger experimental constraints than vector-like lepton states that are uncharged under $U(1)_D$. The possible UV completions of these models are also different. For example, models `a' and `b', could in principle be implemented within a left-right symmetric framework, with both electroweak doublet and singlet BSM fields, while models `c'-`f,' which have vector-like leptons in only one of the two electroweak representations that we observe among the SM leptons, suggest a structure more reminiscent of an $E_6$ grand unified theory.

We see that within this simple framework, our reinforcement learning scan has succeeded both in recreating known models (e.g.~\cite{Wojcik:2022woa}) and identifying models previously not enumerated in the literature. Crucially, this has been accomplished in a framework in which the reinforcement learning agent was required to specify both discrete and continuous particle parameters and identify large hierarchies between different couplings (namely, suppressing the BSM particles' couplings to the electron and $\tau$ while allowing a comparatively large mixing with the muon).

While for a model-building task such as this one, with reasonably finite distinct possible BSM particle content and significant simplifying assumptions that we did not specify to the agent, first among them flavor conservation, the reader may consider the use of machine learning techniques (and the associated demand for computing power) here to be excessive. However, we emphasize again that there is \emph{no} theoretical barrier to applying these same techniques in dramatically more diverse classes of theories, and correspondingly more complex model building tasks. In this light, we can consider the success of the agent in our simple framework as a proof of the underlying concept of our techniques, and leave more ambitious implementations of our procedure to future work.

\section{Summary and Conclusions}
We have presented a method by which reinforcement learning can be used to explore virtually arbitrary spaces of BSM theories, defined only by the theory's symmetry group and the representations that the BSM particles might take, by leveraging the capabilities of graph neural networks. As a case study, we performed an exploration of a simple space of models described by vector-like leptons, inspired by the portal matter framework of \cite{Rizzo:2018vlb,Rizzo:2022lpm}. By evaluating models based on their log-likelihood difference with the SM, the agent was capable of generating both known and novel constructions which addressed the most significant experimental tension with the SM in our input data: The anomalous magnetic moment of the muon. The agent proved capable of generating a significant variety of these constructions, with all six feasible sets of particle content being generated in a single scan. Furthermore, the agent managed to accomplish this task in a somewhat simple, but hardly trivial environment: to identify viable parameter space points, the agent determined both discrete and continuous model parameters, and even learned hierarchical differences between model parameters in order to remain consistent with lepton flavor violation observables.

We emphasize that although our case study is performed for a reasonably simple model, virtually any class of BSM theories might be explored through this technique, without specifying \textit{a priori} the particle content of the model. Therefore, our graph-based approach represents a significant step in automating the process of BSM model building through reinforcement learning. As discussed in our companion work \cite{Wojcik:2024xxx}, the techniques we have outlined here also leave significant possibilities for future refinement and exploration. For example, developing an understanding of the scaling of this procedure's efficacy with more elaborate model spaces requires further experimentation, and various significant refinements, such as implementing scans of continuous parameter spaces with Monte Carlo techniques once the reinforcement agent has specified parameters' orders of magnitude, might radically improve the methodology's efficiency.

Given its generality, the development of this procedure into an automated model-building tool is a significant, but feasible, undertaking. In particular, if the generation of graph grammars for different classes of models can be automated, the large variety of automatic computing tools for BSM observables \cite{Straub:2018kue,Uhlrich:2020ltd,Belanger:2010pz,Alwall:2014hca,Stelzer:1994ta} could be readily incorporated into a pipeline which rapidly implements an environment and performs these reinforcement learning scans.  

In our companion paper \cite{Wojcik:2024xxx}, we also highlight the utility of mathematical graphs in representing BSM models for a variety of machine learning tasks beyond the reinforcement learning scan represented here, and enumerate some further possibilities, such as developing a BSM model builder based on generative AI techniques rather than reinforcement learning. Given the demonstrated capabilities of graph neural networks in other contexts, the applications of this technology to the study of the space of BSM models are potentially significant and merit substantial further exploration.

\section*{Acknowledgements}
This work was supported by the U.S. Department of Energy under the contract number DE-SC0017647.

\bibliography{main}

\begin{thebibliography}{81}%
\makeatletter
\providecommand \@ifxundefined [1]{%
 \@ifx{#1\undefined}
}%
\providecommand \@ifnum [1]{%
 \ifnum #1\expandafter \@firstoftwo
 \else \expandafter \@secondoftwo
 \fi
}%
\providecommand \@ifx [1]{%
 \ifx #1\expandafter \@firstoftwo
 \else \expandafter \@secondoftwo
 \fi
}%
\providecommand \natexlab [1]{#1}%
\providecommand \enquote  [1]{``#1''}%
\providecommand \bibnamefont  [1]{#1}%
\providecommand \bibfnamefont [1]{#1}%
\providecommand \citenamefont [1]{#1}%
\providecommand \href@noop [0]{\@secondoftwo}%
\providecommand \href [0]{\begingroup \@sanitize@url \@href}%
\providecommand \@href[1]{\@@startlink{#1}\@@href}%
\providecommand \@@href[1]{\endgroup#1\@@endlink}%
\providecommand \@sanitize@url [0]{\catcode `\\12\catcode `\$12\catcode `\&12\catcode `\#12\catcode `\^12\catcode `\_12\catcode `\%12\relax}%
\providecommand \@@startlink[1]{}%
\providecommand \@@endlink[0]{}%
\providecommand \url  [0]{\begingroup\@sanitize@url \@url }%
\providecommand \@url [1]{\endgroup\@href {#1}{\urlprefix }}%
\providecommand \urlprefix  [0]{URL }%
\providecommand \Eprint [0]{\href }%
\providecommand \doibase [0]{https://doi.org/}%
\providecommand \selectlanguage [0]{\@gobble}%
\providecommand \bibinfo  [0]{\@secondoftwo}%
\providecommand \bibfield  [0]{\@secondoftwo}%
\providecommand \translation [1]{[#1]}%
\providecommand \BibitemOpen [0]{}%
\providecommand \bibitemStop [0]{}%
\providecommand \bibitemNoStop [0]{.\EOS\space}%
\providecommand \EOS [0]{\spacefactor3000\relax}%
\providecommand \BibitemShut  [1]{\csname bibitem#1\endcsname}%
\let\auto@bib@innerbib\@empty
\bibitem [{\citenamefont {Halverson}\ \emph {et~al.}(2019)\citenamefont {Halverson}, \citenamefont {Nelson},\ and\ \citenamefont {Ruehle}}]{Halverson:2019tkf}%
  \BibitemOpen
  \bibfield  {author} {\bibinfo {author} {\bibfnamefont {J.}~\bibnamefont {Halverson}}, \bibinfo {author} {\bibfnamefont {B.}~\bibnamefont {Nelson}},\ and\ \bibinfo {author} {\bibfnamefont {F.}~\bibnamefont {Ruehle}},\ }\bibfield  {title} {\bibinfo {title} {{Branes with Brains: Exploring String Vacua with Deep Reinforcement Learning}},\ }\href {https://doi.org/10.1007/JHEP06(2019)003} {\bibfield  {journal} {\bibinfo  {journal} {JHEP}\ }\textbf {\bibinfo {volume} {06}},\ \bibinfo {pages} {003}},\ \Eprint {https://arxiv.org/abs/1903.11616} {arXiv:1903.11616 [hep-th]} \BibitemShut {NoStop}%
\bibitem [{\citenamefont {Harvey}\ and\ \citenamefont {Lukas}(2021)}]{Harvey:2021oue}%
  \BibitemOpen
  \bibfield  {author} {\bibinfo {author} {\bibfnamefont {T.~R.}\ \bibnamefont {Harvey}}\ and\ \bibinfo {author} {\bibfnamefont {A.}~\bibnamefont {Lukas}},\ }\bibfield  {title} {\bibinfo {title} {{Quark Mass Models and Reinforcement Learning}},\ }\href {https://doi.org/10.1007/JHEP08(2021)161} {\bibfield  {journal} {\bibinfo  {journal} {JHEP}\ }\textbf {\bibinfo {volume} {08}},\ \bibinfo {pages} {161}},\ \Eprint {https://arxiv.org/abs/2103.04759} {arXiv:2103.04759 [hep-th]} \BibitemShut {NoStop}%
\bibitem [{\citenamefont {Nishimura}\ \emph {et~al.}(2020)\citenamefont {Nishimura}, \citenamefont {Miyao},\ and\ \citenamefont {Otsuka}}]{Nishimura:2020nre}%
  \BibitemOpen
  \bibfield  {author} {\bibinfo {author} {\bibfnamefont {S.}~\bibnamefont {Nishimura}}, \bibinfo {author} {\bibfnamefont {C.}~\bibnamefont {Miyao}},\ and\ \bibinfo {author} {\bibfnamefont {H.}~\bibnamefont {Otsuka}},\ }\bibfield  {title} {\bibinfo {title} {{Exploring the flavor structure of quarks and leptons with reinforcement learning}},\ }\href {https://doi.org/10.1007/JHEP12(2023)021} {\bibfield  {journal} {\bibinfo  {journal} {JHEP}\ }\textbf {\bibinfo {volume} {23}},\ \bibinfo {pages} {021}},\ \Eprint {https://arxiv.org/abs/2304.14176} {arXiv:2304.14176 [hep-ph]} \BibitemShut {NoStop}%
\bibitem [{\citenamefont {Rizzo}(2019)}]{Rizzo:2018vlb}%
  \BibitemOpen
  \bibfield  {author} {\bibinfo {author} {\bibfnamefont {T.~G.}\ \bibnamefont {Rizzo}},\ }\bibfield  {title} {\bibinfo {title} {{Kinetic Mixing and Portal Matter Phenomenology}},\ }\href {https://doi.org/10.1103/PhysRevD.99.115024} {\bibfield  {journal} {\bibinfo  {journal} {Phys. Rev. D}\ }\textbf {\bibinfo {volume} {99}},\ \bibinfo {pages} {115024} (\bibinfo {year} {2019})},\ \Eprint {https://arxiv.org/abs/1810.07531} {arXiv:1810.07531 [hep-ph]} \BibitemShut {NoStop}%
\bibitem [{\citenamefont {Rizzo}(2022)}]{Rizzo:2022lpm}%
  \BibitemOpen
  \bibfield  {author} {\bibinfo {author} {\bibfnamefont {T.~G.}\ \bibnamefont {Rizzo}},\ }\bibfield  {title} {\bibinfo {title} {{Toward a UV model of kinetic mixing and portal matter. II. Exploring unification in an SU(N) group}},\ }\href {https://doi.org/10.1103/PhysRevD.106.095024} {\bibfield  {journal} {\bibinfo  {journal} {Phys. Rev. D}\ }\textbf {\bibinfo {volume} {106}},\ \bibinfo {pages} {095024} (\bibinfo {year} {2022})},\ \Eprint {https://arxiv.org/abs/2209.00688} {arXiv:2209.00688 [hep-ph]} \BibitemShut {NoStop}%
\bibitem [{\citenamefont {Bennett}\ \emph {et~al.}(2006)\citenamefont {Bennett} \emph {et~al.}}]{Muong-2:2006rrc}%
  \BibitemOpen
  \bibfield  {author} {\bibinfo {author} {\bibfnamefont {G.~W.}\ \bibnamefont {Bennett}} \emph {et~al.} (\bibinfo {collaboration} {Muon g-2}),\ }\bibfield  {title} {\bibinfo {title} {{Final Report of the Muon E821 Anomalous Magnetic Moment Measurement at BNL}},\ }\href {https://doi.org/10.1103/PhysRevD.73.072003} {\bibfield  {journal} {\bibinfo  {journal} {Phys. Rev. D}\ }\textbf {\bibinfo {volume} {73}},\ \bibinfo {pages} {072003} (\bibinfo {year} {2006})},\ \Eprint {https://arxiv.org/abs/hep-ex/0602035} {arXiv:hep-ex/0602035} \BibitemShut {NoStop}%
\bibitem [{\citenamefont {Abi}\ \emph {et~al.}(2021)\citenamefont {Abi} \emph {et~al.}}]{Muong-2:2021ojo}%
  \BibitemOpen
  \bibfield  {author} {\bibinfo {author} {\bibfnamefont {B.}~\bibnamefont {Abi}} \emph {et~al.} (\bibinfo {collaboration} {Muon g-2}),\ }\bibfield  {title} {\bibinfo {title} {{Measurement of the Positive Muon Anomalous Magnetic Moment to 0.46 ppm}},\ }\href {https://doi.org/10.1103/PhysRevLett.126.141801} {\bibfield  {journal} {\bibinfo  {journal} {Phys. Rev. Lett.}\ }\textbf {\bibinfo {volume} {126}},\ \bibinfo {pages} {141801} (\bibinfo {year} {2021})},\ \Eprint {https://arxiv.org/abs/2104.03281} {arXiv:2104.03281 [hep-ex]} \BibitemShut {NoStop}%
\bibitem [{\citenamefont {Aguillard}\ \emph {et~al.}(2023)\citenamefont {Aguillard} \emph {et~al.}}]{Muong-2:2023cdq}%
  \BibitemOpen
  \bibfield  {author} {\bibinfo {author} {\bibfnamefont {D.~P.}\ \bibnamefont {Aguillard}} \emph {et~al.} (\bibinfo {collaboration} {Muon g-2}),\ }\bibfield  {title} {\bibinfo {title} {{Measurement of the Positive Muon Anomalous Magnetic Moment to 0.20~ppm}},\ }\href {https://doi.org/10.1103/PhysRevLett.131.161802} {\bibfield  {journal} {\bibinfo  {journal} {Phys. Rev. Lett.}\ }\textbf {\bibinfo {volume} {131}},\ \bibinfo {pages} {161802} (\bibinfo {year} {2023})},\ \Eprint {https://arxiv.org/abs/2308.06230} {arXiv:2308.06230 [hep-ex]} \BibitemShut {NoStop}%
\bibitem [{\citenamefont {Aoyama}\ \emph {et~al.}(2020)\citenamefont {Aoyama} \emph {et~al.}}]{Aoyama:2020ynm}%
  \BibitemOpen
  \bibfield  {author} {\bibinfo {author} {\bibfnamefont {T.}~\bibnamefont {Aoyama}} \emph {et~al.},\ }\bibfield  {title} {\bibinfo {title} {{The anomalous magnetic moment of the muon in the Standard Model}},\ }\href {https://doi.org/10.1016/j.physrep.2020.07.006} {\bibfield  {journal} {\bibinfo  {journal} {Phys. Rept.}\ }\textbf {\bibinfo {volume} {887}},\ \bibinfo {pages} {1} (\bibinfo {year} {2020})},\ \Eprint {https://arxiv.org/abs/2006.04822} {arXiv:2006.04822 [hep-ph]} \BibitemShut {NoStop}%
\bibitem [{\citenamefont {Aoyama}\ \emph {et~al.}(2012)\citenamefont {Aoyama}, \citenamefont {Hayakawa}, \citenamefont {Kinoshita},\ and\ \citenamefont {Nio}}]{Aoyama:2012wk}%
  \BibitemOpen
  \bibfield  {author} {\bibinfo {author} {\bibfnamefont {T.}~\bibnamefont {Aoyama}}, \bibinfo {author} {\bibfnamefont {M.}~\bibnamefont {Hayakawa}}, \bibinfo {author} {\bibfnamefont {T.}~\bibnamefont {Kinoshita}},\ and\ \bibinfo {author} {\bibfnamefont {M.}~\bibnamefont {Nio}},\ }\bibfield  {title} {\bibinfo {title} {{Complete Tenth-Order QED Contribution to the Muon $g-2$}},\ }\href {https://doi.org/10.1103/PhysRevLett.109.111808} {\bibfield  {journal} {\bibinfo  {journal} {Phys. Rev. Lett.}\ }\textbf {\bibinfo {volume} {109}},\ \bibinfo {pages} {111808} (\bibinfo {year} {2012})},\ \Eprint {https://arxiv.org/abs/1205.5370} {arXiv:1205.5370 [hep-ph]} \BibitemShut {NoStop}%
\bibitem [{\citenamefont {Aoyama}\ \emph {et~al.}(2019)\citenamefont {Aoyama}, \citenamefont {Kinoshita},\ and\ \citenamefont {Nio}}]{Aoyama:2019ryr}%
  \BibitemOpen
  \bibfield  {author} {\bibinfo {author} {\bibfnamefont {T.}~\bibnamefont {Aoyama}}, \bibinfo {author} {\bibfnamefont {T.}~\bibnamefont {Kinoshita}},\ and\ \bibinfo {author} {\bibfnamefont {M.}~\bibnamefont {Nio}},\ }\bibfield  {title} {\bibinfo {title} {{Theory of the Anomalous Magnetic Moment of the Electron}},\ }\href {https://doi.org/10.3390/atoms7010028} {\bibfield  {journal} {\bibinfo  {journal} {Atoms}\ }\textbf {\bibinfo {volume} {7}},\ \bibinfo {pages} {28} (\bibinfo {year} {2019})}\BibitemShut {NoStop}%
\bibitem [{\citenamefont {Czarnecki}\ \emph {et~al.}(2003)\citenamefont {Czarnecki}, \citenamefont {Marciano},\ and\ \citenamefont {Vainshtein}}]{Czarnecki:2002nt}%
  \BibitemOpen
  \bibfield  {author} {\bibinfo {author} {\bibfnamefont {A.}~\bibnamefont {Czarnecki}}, \bibinfo {author} {\bibfnamefont {W.~J.}\ \bibnamefont {Marciano}},\ and\ \bibinfo {author} {\bibfnamefont {A.}~\bibnamefont {Vainshtein}},\ }\bibfield  {title} {\bibinfo {title} {{Refinements in electroweak contributions to the muon anomalous magnetic moment}},\ }\href {https://doi.org/10.1103/PhysRevD.67.073006} {\bibfield  {journal} {\bibinfo  {journal} {Phys. Rev.}\ }\textbf {\bibinfo {volume} {D67}},\ \bibinfo {pages} {073006} (\bibinfo {year} {2003})},\ \bibinfo {note} {[Erratum: Phys. Rev. {\bf D73}, 119901 (2006)]},\ \Eprint {https://arxiv.org/abs/hep-ph/0212229} {arXiv:hep-ph/0212229 [hep-ph]} \BibitemShut {NoStop}%
\bibitem [{\citenamefont {Gnendiger}\ \emph {et~al.}(2013)\citenamefont {Gnendiger}, \citenamefont {St{\"o}ckinger},\ and\ \citenamefont {St{\"o}ckinger-Kim}}]{Gnendiger:2013pva}%
  \BibitemOpen
  \bibfield  {author} {\bibinfo {author} {\bibfnamefont {C.}~\bibnamefont {Gnendiger}}, \bibinfo {author} {\bibfnamefont {D.}~\bibnamefont {St{\"o}ckinger}},\ and\ \bibinfo {author} {\bibfnamefont {H.}~\bibnamefont {St{\"o}ckinger-Kim}},\ }\bibfield  {title} {\bibinfo {title} {{The electroweak contributions to $(g-2)_\mu$ after the Higgs boson mass measurement}},\ }\href {https://doi.org/10.1103/PhysRevD.88.053005} {\bibfield  {journal} {\bibinfo  {journal} {Phys. Rev.}\ }\textbf {\bibinfo {volume} {D88}},\ \bibinfo {pages} {053005} (\bibinfo {year} {2013})},\ \Eprint {https://arxiv.org/abs/1306.5546} {arXiv:1306.5546 [hep-ph]} \BibitemShut {NoStop}%
\bibitem [{\citenamefont {Davier}\ \emph {et~al.}(2017)\citenamefont {Davier}, \citenamefont {Hoecker}, \citenamefont {Malaescu},\ and\ \citenamefont {Zhang}}]{Davier:2017zfy}%
  \BibitemOpen
  \bibfield  {author} {\bibinfo {author} {\bibfnamefont {M.}~\bibnamefont {Davier}}, \bibinfo {author} {\bibfnamefont {A.}~\bibnamefont {Hoecker}}, \bibinfo {author} {\bibfnamefont {B.}~\bibnamefont {Malaescu}},\ and\ \bibinfo {author} {\bibfnamefont {Z.}~\bibnamefont {Zhang}},\ }\bibfield  {title} {\bibinfo {title} {{Reevaluation of the hadronic vacuum polarisation contributions to the Standard Model predictions of the muon $g-2$ and ${\alpha (m_Z^2)}$ using newest hadronic cross-section data}},\ }\href {https://doi.org/10.1140/epjc/s10052-017-5161-6} {\bibfield  {journal} {\bibinfo  {journal} {Eur. Phys. J.}\ }\textbf {\bibinfo {volume} {C77}},\ \bibinfo {pages} {827} (\bibinfo {year} {2017})},\ \Eprint {https://arxiv.org/abs/1706.09436} {arXiv:1706.09436 [hep-ph]} \BibitemShut {NoStop}%
\bibitem [{\citenamefont {Keshavarzi}\ \emph {et~al.}(2018)\citenamefont {Keshavarzi}, \citenamefont {Nomura},\ and\ \citenamefont {Teubner}}]{Keshavarzi:2018mgv}%
  \BibitemOpen
  \bibfield  {author} {\bibinfo {author} {\bibfnamefont {A.}~\bibnamefont {Keshavarzi}}, \bibinfo {author} {\bibfnamefont {D.}~\bibnamefont {Nomura}},\ and\ \bibinfo {author} {\bibfnamefont {T.}~\bibnamefont {Teubner}},\ }\bibfield  {title} {\bibinfo {title} {{Muon $g-2$ and $\alpha(M_Z^2)$: a new data-based analysis}},\ }\href {https://doi.org/10.1103/PhysRevD.97.114025} {\bibfield  {journal} {\bibinfo  {journal} {Phys. Rev.}\ }\textbf {\bibinfo {volume} {D97}},\ \bibinfo {pages} {114025} (\bibinfo {year} {2018})},\ \Eprint {https://arxiv.org/abs/1802.02995} {arXiv:1802.02995 [hep-ph]} \BibitemShut {NoStop}%
\bibitem [{\citenamefont {Colangelo}\ \emph {et~al.}(2019)\citenamefont {Colangelo}, \citenamefont {Hoferichter},\ and\ \citenamefont {Stoffer}}]{Colangelo:2018mtw}%
  \BibitemOpen
  \bibfield  {author} {\bibinfo {author} {\bibfnamefont {G.}~\bibnamefont {Colangelo}}, \bibinfo {author} {\bibfnamefont {M.}~\bibnamefont {Hoferichter}},\ and\ \bibinfo {author} {\bibfnamefont {P.}~\bibnamefont {Stoffer}},\ }\bibfield  {title} {\bibinfo {title} {{Two-pion contribution to hadronic vacuum polarization}},\ }\href {https://doi.org/10.1007/JHEP02(2019)006} {\bibfield  {journal} {\bibinfo  {journal} {JHEP}\ }\textbf {\bibinfo {volume} {02}},\ \bibinfo {pages} {006}},\ \Eprint {https://arxiv.org/abs/1810.00007} {arXiv:1810.00007 [hep-ph]} \BibitemShut {NoStop}%
\bibitem [{\citenamefont {Hoferichter}\ \emph {et~al.}(2019)\citenamefont {Hoferichter}, \citenamefont {Hoid},\ and\ \citenamefont {Kubis}}]{Hoferichter:2019mqg}%
  \BibitemOpen
  \bibfield  {author} {\bibinfo {author} {\bibfnamefont {M.}~\bibnamefont {Hoferichter}}, \bibinfo {author} {\bibfnamefont {B.-L.}\ \bibnamefont {Hoid}},\ and\ \bibinfo {author} {\bibfnamefont {B.}~\bibnamefont {Kubis}},\ }\bibfield  {title} {\bibinfo {title} {{Three-pion contribution to hadronic vacuum polarization}},\ }\href {https://doi.org/10.1007/JHEP08(2019)137} {\bibfield  {journal} {\bibinfo  {journal} {JHEP}\ }\textbf {\bibinfo {volume} {08}},\ \bibinfo {pages} {137}},\ \Eprint {https://arxiv.org/abs/1907.01556} {arXiv:1907.01556 [hep-ph]} \BibitemShut {NoStop}%
\bibitem [{\citenamefont {Davier}\ \emph {et~al.}(2020)\citenamefont {Davier}, \citenamefont {Hoecker}, \citenamefont {Malaescu},\ and\ \citenamefont {Zhang}}]{Davier:2019can}%
  \BibitemOpen
  \bibfield  {author} {\bibinfo {author} {\bibfnamefont {M.}~\bibnamefont {Davier}}, \bibinfo {author} {\bibfnamefont {A.}~\bibnamefont {Hoecker}}, \bibinfo {author} {\bibfnamefont {B.}~\bibnamefont {Malaescu}},\ and\ \bibinfo {author} {\bibfnamefont {Z.}~\bibnamefont {Zhang}},\ }\bibfield  {title} {\bibinfo {title} {{A new evaluation of the hadronic vacuum polarisation contributions to the muon anomalous magnetic moment and to $\mathbf{\boldsymbol\alpha(m_Z^2)}$}},\ }\href {https://doi.org/10.1140/epjc/s10052-020-7792-2} {\bibfield  {journal} {\bibinfo  {journal} {Eur. Phys. J.}\ }\textbf {\bibinfo {volume} {C80}},\ \bibinfo {pages} {241} (\bibinfo {year} {2020})},\ \bibinfo {note} {[Erratum: Eur. Phys. J. {\bf C80}, 410 (2020)]},\ \Eprint {https://arxiv.org/abs/1908.00921} {arXiv:1908.00921 [hep-ph]} \BibitemShut {NoStop}%
\bibitem [{\citenamefont {Keshavarzi}\ \emph {et~al.}(2020)\citenamefont {Keshavarzi}, \citenamefont {Nomura},\ and\ \citenamefont {Teubner}}]{Keshavarzi:2019abf}%
  \BibitemOpen
  \bibfield  {author} {\bibinfo {author} {\bibfnamefont {A.}~\bibnamefont {Keshavarzi}}, \bibinfo {author} {\bibfnamefont {D.}~\bibnamefont {Nomura}},\ and\ \bibinfo {author} {\bibfnamefont {T.}~\bibnamefont {Teubner}},\ }\bibfield  {title} {\bibinfo {title} {{The $g-2$ of charged leptons, $\alpha(M_Z^2)$ and the hyperfine splitting of muonium}},\ }\href {https://doi.org/10.1103/PhysRevD.101.014029} {\bibfield  {journal} {\bibinfo  {journal} {Phys. Rev.}\ }\textbf {\bibinfo {volume} {D101}},\ \bibinfo {pages} {014029} (\bibinfo {year} {2020})},\ \Eprint {https://arxiv.org/abs/1911.00367} {arXiv:1911.00367 [hep-ph]} \BibitemShut {NoStop}%
\bibitem [{\citenamefont {Kurz}\ \emph {et~al.}(2014)\citenamefont {Kurz}, \citenamefont {Liu}, \citenamefont {Marquard},\ and\ \citenamefont {Steinhauser}}]{Kurz:2014wya}%
  \BibitemOpen
  \bibfield  {author} {\bibinfo {author} {\bibfnamefont {A.}~\bibnamefont {Kurz}}, \bibinfo {author} {\bibfnamefont {T.}~\bibnamefont {Liu}}, \bibinfo {author} {\bibfnamefont {P.}~\bibnamefont {Marquard}},\ and\ \bibinfo {author} {\bibfnamefont {M.}~\bibnamefont {Steinhauser}},\ }\bibfield  {title} {\bibinfo {title} {{Hadronic contribution to the muon anomalous magnetic moment to next-to-next-to-leading order}},\ }\href {https://doi.org/10.1016/j.physletb.2014.05.043} {\bibfield  {journal} {\bibinfo  {journal} {Phys. Lett.}\ }\textbf {\bibinfo {volume} {B734}},\ \bibinfo {pages} {144} (\bibinfo {year} {2014})},\ \Eprint {https://arxiv.org/abs/1403.6400} {arXiv:1403.6400 [hep-ph]} \BibitemShut {NoStop}%
\bibitem [{\citenamefont {Melnikov}\ and\ \citenamefont {Vainshtein}(2004)}]{Melnikov:2003xd}%
  \BibitemOpen
  \bibfield  {author} {\bibinfo {author} {\bibfnamefont {K.}~\bibnamefont {Melnikov}}\ and\ \bibinfo {author} {\bibfnamefont {A.}~\bibnamefont {Vainshtein}},\ }\bibfield  {title} {\bibinfo {title} {{Hadronic light-by-light scattering contribution to the muon anomalous magnetic moment revisited}},\ }\href {https://doi.org/10.1103/PhysRevD.70.113006} {\bibfield  {journal} {\bibinfo  {journal} {Phys. Rev.}\ }\textbf {\bibinfo {volume} {D70}},\ \bibinfo {pages} {113006} (\bibinfo {year} {2004})},\ \Eprint {https://arxiv.org/abs/hep-ph/0312226} {arXiv:hep-ph/0312226 [hep-ph]} \BibitemShut {NoStop}%
\bibitem [{\citenamefont {Masjuan}\ and\ \citenamefont {S{\'a}nchez-Puertas}(2017)}]{Masjuan:2017tvw}%
  \BibitemOpen
  \bibfield  {author} {\bibinfo {author} {\bibfnamefont {P.}~\bibnamefont {Masjuan}}\ and\ \bibinfo {author} {\bibfnamefont {P.}~\bibnamefont {S{\'a}nchez-Puertas}},\ }\bibfield  {title} {\bibinfo {title} {{Pseudoscalar-pole contribution to the $(g_{\mu}-2)$: a rational approach}},\ }\href {https://doi.org/10.1103/PhysRevD.95.054026} {\bibfield  {journal} {\bibinfo  {journal} {Phys. Rev.}\ }\textbf {\bibinfo {volume} {D95}},\ \bibinfo {pages} {054026} (\bibinfo {year} {2017})},\ \Eprint {https://arxiv.org/abs/1701.05829} {arXiv:1701.05829 [hep-ph]} \BibitemShut {NoStop}%
\bibitem [{\citenamefont {Colangelo}\ \emph {et~al.}(2017)\citenamefont {Colangelo}, \citenamefont {Hoferichter}, \citenamefont {Procura},\ and\ \citenamefont {Stoffer}}]{Colangelo:2017fiz}%
  \BibitemOpen
  \bibfield  {author} {\bibinfo {author} {\bibfnamefont {G.}~\bibnamefont {Colangelo}}, \bibinfo {author} {\bibfnamefont {M.}~\bibnamefont {Hoferichter}}, \bibinfo {author} {\bibfnamefont {M.}~\bibnamefont {Procura}},\ and\ \bibinfo {author} {\bibfnamefont {P.}~\bibnamefont {Stoffer}},\ }\bibfield  {title} {\bibinfo {title} {{Dispersion relation for hadronic light-by-light scattering: two-pion contributions}},\ }\href {https://doi.org/10.1007/JHEP04(2017)161} {\bibfield  {journal} {\bibinfo  {journal} {JHEP}\ }\textbf {\bibinfo {volume} {04}},\ \bibinfo {pages} {161}},\ \Eprint {https://arxiv.org/abs/1702.07347} {arXiv:1702.07347 [hep-ph]} \BibitemShut {NoStop}%
\bibitem [{\citenamefont {Hoferichter}\ \emph {et~al.}(2018)\citenamefont {Hoferichter}, \citenamefont {Hoid}, \citenamefont {Kubis}, \citenamefont {Leupold},\ and\ \citenamefont {Schneider}}]{Hoferichter:2018kwz}%
  \BibitemOpen
  \bibfield  {author} {\bibinfo {author} {\bibfnamefont {M.}~\bibnamefont {Hoferichter}}, \bibinfo {author} {\bibfnamefont {B.-L.}\ \bibnamefont {Hoid}}, \bibinfo {author} {\bibfnamefont {B.}~\bibnamefont {Kubis}}, \bibinfo {author} {\bibfnamefont {S.}~\bibnamefont {Leupold}},\ and\ \bibinfo {author} {\bibfnamefont {S.~P.}\ \bibnamefont {Schneider}},\ }\bibfield  {title} {\bibinfo {title} {{Dispersion relation for hadronic light-by-light scattering: pion pole}},\ }\href {https://doi.org/10.1007/JHEP10(2018)141} {\bibfield  {journal} {\bibinfo  {journal} {JHEP}\ }\textbf {\bibinfo {volume} {10}},\ \bibinfo {pages} {141}},\ \Eprint {https://arxiv.org/abs/1808.04823} {arXiv:1808.04823 [hep-ph]} \BibitemShut {NoStop}%
\bibitem [{\citenamefont {G{\'e}rardin}\ \emph {et~al.}(2019)\citenamefont {G{\'e}rardin}, \citenamefont {Meyer},\ and\ \citenamefont {Nyffeler}}]{Gerardin:2019vio}%
  \BibitemOpen
  \bibfield  {author} {\bibinfo {author} {\bibfnamefont {A.}~\bibnamefont {G{\'e}rardin}}, \bibinfo {author} {\bibfnamefont {H.~B.}\ \bibnamefont {Meyer}},\ and\ \bibinfo {author} {\bibfnamefont {A.}~\bibnamefont {Nyffeler}},\ }\bibfield  {title} {\bibinfo {title} {{Lattice calculation of the pion transition form factor with $N_f=2+1$ Wilson quarks}},\ }\href {https://doi.org/10.1103/PhysRevD.100.034520} {\bibfield  {journal} {\bibinfo  {journal} {Phys. Rev.}\ }\textbf {\bibinfo {volume} {D100}},\ \bibinfo {pages} {034520} (\bibinfo {year} {2019})},\ \Eprint {https://arxiv.org/abs/1903.09471} {arXiv:1903.09471 [hep-lat]} \BibitemShut {NoStop}%
\bibitem [{\citenamefont {Bijnens}\ \emph {et~al.}(2019)\citenamefont {Bijnens}, \citenamefont {Hermansson-Truedsson},\ and\ \citenamefont {Rodr{\'i}guez-S{\'a}nchez}}]{Bijnens:2019ghy}%
  \BibitemOpen
  \bibfield  {author} {\bibinfo {author} {\bibfnamefont {J.}~\bibnamefont {Bijnens}}, \bibinfo {author} {\bibfnamefont {N.}~\bibnamefont {Hermansson-Truedsson}},\ and\ \bibinfo {author} {\bibfnamefont {A.}~\bibnamefont {Rodr{\'i}guez-S{\'a}nchez}},\ }\bibfield  {title} {\bibinfo {title} {{Short-distance constraints for the HLbL contribution to the muon anomalous magnetic moment}},\ }\href {https://doi.org/10.1016/j.physletb.2019.134994} {\bibfield  {journal} {\bibinfo  {journal} {Phys. Lett.}\ }\textbf {\bibinfo {volume} {B798}},\ \bibinfo {pages} {134994} (\bibinfo {year} {2019})},\ \Eprint {https://arxiv.org/abs/1908.03331} {arXiv:1908.03331 [hep-ph]} \BibitemShut {NoStop}%
\bibitem [{\citenamefont {Colangelo}\ \emph {et~al.}(2020)\citenamefont {Colangelo}, \citenamefont {Hagelstein}, \citenamefont {Hoferichter}, \citenamefont {Laub},\ and\ \citenamefont {Stoffer}}]{Colangelo:2019uex}%
  \BibitemOpen
  \bibfield  {author} {\bibinfo {author} {\bibfnamefont {G.}~\bibnamefont {Colangelo}}, \bibinfo {author} {\bibfnamefont {F.}~\bibnamefont {Hagelstein}}, \bibinfo {author} {\bibfnamefont {M.}~\bibnamefont {Hoferichter}}, \bibinfo {author} {\bibfnamefont {L.}~\bibnamefont {Laub}},\ and\ \bibinfo {author} {\bibfnamefont {P.}~\bibnamefont {Stoffer}},\ }\bibfield  {title} {\bibinfo {title} {{Longitudinal short-distance constraints for the hadronic light-by-light contribution to $(g-2)_\mu$ with large-$N_c$ Regge models}},\ }\href {https://doi.org/10.1007/JHEP03(2020)101} {\bibfield  {journal} {\bibinfo  {journal} {JHEP}\ }\textbf {\bibinfo {volume} {03}},\ \bibinfo {pages} {101}},\ \Eprint {https://arxiv.org/abs/1910.13432} {arXiv:1910.13432 [hep-ph]} \BibitemShut {NoStop}%
\bibitem [{\citenamefont {Blum}\ \emph {et~al.}(2020)\citenamefont {Blum}, \citenamefont {Christ}, \citenamefont {Hayakawa}, \citenamefont {Izubuchi}, \citenamefont {Jin}, \citenamefont {Jung},\ and\ \citenamefont {Lehner}}]{Blum:2019ugy}%
  \BibitemOpen
  \bibfield  {author} {\bibinfo {author} {\bibfnamefont {T.}~\bibnamefont {Blum}}, \bibinfo {author} {\bibfnamefont {N.}~\bibnamefont {Christ}}, \bibinfo {author} {\bibfnamefont {M.}~\bibnamefont {Hayakawa}}, \bibinfo {author} {\bibfnamefont {T.}~\bibnamefont {Izubuchi}}, \bibinfo {author} {\bibfnamefont {L.}~\bibnamefont {Jin}}, \bibinfo {author} {\bibfnamefont {C.}~\bibnamefont {Jung}},\ and\ \bibinfo {author} {\bibfnamefont {C.}~\bibnamefont {Lehner}},\ }\bibfield  {title} {\bibinfo {title} {{The hadronic light-by-light scattering contribution to the muon anomalous magnetic moment from lattice QCD}},\ }\href {https://doi.org/10.1103/PhysRevLett.124.132002} {\bibfield  {journal} {\bibinfo  {journal} {Phys. Rev. Lett.}\ }\textbf {\bibinfo {volume} {124}},\ \bibinfo {pages} {132002} (\bibinfo {year} {2020})},\ \Eprint {https://arxiv.org/abs/1911.08123} {arXiv:1911.08123 [hep-lat]} \BibitemShut {NoStop}%
\bibitem [{\citenamefont {Colangelo}\ \emph {et~al.}(2014)\citenamefont {Colangelo}, \citenamefont {Hoferichter}, \citenamefont {Nyffeler}, \citenamefont {Passera},\ and\ \citenamefont {Stoffer}}]{Colangelo:2014qya}%
  \BibitemOpen
  \bibfield  {author} {\bibinfo {author} {\bibfnamefont {G.}~\bibnamefont {Colangelo}}, \bibinfo {author} {\bibfnamefont {M.}~\bibnamefont {Hoferichter}}, \bibinfo {author} {\bibfnamefont {A.}~\bibnamefont {Nyffeler}}, \bibinfo {author} {\bibfnamefont {M.}~\bibnamefont {Passera}},\ and\ \bibinfo {author} {\bibfnamefont {P.}~\bibnamefont {Stoffer}},\ }\bibfield  {title} {\bibinfo {title} {{Remarks on higher-order hadronic corrections to the muon $g-2$}},\ }\href {https://doi.org/10.1016/j.physletb.2014.06.012} {\bibfield  {journal} {\bibinfo  {journal} {Phys. Lett.}\ }\textbf {\bibinfo {volume} {B735}},\ \bibinfo {pages} {90} (\bibinfo {year} {2014})},\ \Eprint {https://arxiv.org/abs/1403.7512} {arXiv:1403.7512 [hep-ph]} \BibitemShut {NoStop}%
\bibitem [{\citenamefont {Wojcik}\ \emph {et~al.}(2023)\citenamefont {Wojcik}, \citenamefont {Everett}, \citenamefont {Eu},\ and\ \citenamefont {Ximenes}}]{Wojcik:2022woa}%
  \BibitemOpen
  \bibfield  {author} {\bibinfo {author} {\bibfnamefont {G.~N.}\ \bibnamefont {Wojcik}}, \bibinfo {author} {\bibfnamefont {L.~L.}\ \bibnamefont {Everett}}, \bibinfo {author} {\bibfnamefont {S.~T.}\ \bibnamefont {Eu}},\ and\ \bibinfo {author} {\bibfnamefont {R.}~\bibnamefont {Ximenes}},\ }\bibfield  {title} {\bibinfo {title} {{Portal matter, kinetic mixing, and muon g\ensuremath{-}2}},\ }\href {https://doi.org/10.1016/j.physletb.2023.137931} {\bibfield  {journal} {\bibinfo  {journal} {Phys. Lett. B}\ }\textbf {\bibinfo {volume} {841}},\ \bibinfo {pages} {137931} (\bibinfo {year} {2023})},\ \Eprint {https://arxiv.org/abs/2211.09918} {arXiv:2211.09918 [hep-ph]} \BibitemShut {NoStop}%
\bibitem [{\citenamefont {Wojcik}\ \emph {et~al.}()\citenamefont {Wojcik}, \citenamefont {Eu},\ and\ \citenamefont {Everett}}]{Wojcik:2024xxx}%
  \BibitemOpen
  \bibfield  {author} {\bibinfo {author} {\bibfnamefont {G.~N.}\ \bibnamefont {Wojcik}}, \bibinfo {author} {\bibfnamefont {S.~T.}\ \bibnamefont {Eu}},\ and\ \bibinfo {author} {\bibfnamefont {L.~L.}\ \bibnamefont {Everett}},\ }\bibfield  {title} {\bibinfo {title} {Graph reinforcement learning for exploring beyond-standard-model model spaces},\ }\href@noop {} {\bibinfo  {journal} {{to appear}}\ }\BibitemShut {NoStop}%
\bibitem [{Note1()}]{Note1}%
  \BibitemOpen
\bibfield  {journal} {  }\bibinfo {note} {It is in principle possible to express an identical graph grammar using a traditional homogeneous graph, without differing node types. For more details on the subtleties of this construction we refer the reader to \cite {Wojcik:2024xxx}.}\BibitemShut {Stop}%
\bibitem [{\citenamefont {Workman}\ and\ \citenamefont {Others}(2022)}]{Workman:2022ynf}%
  \BibitemOpen
  \bibfield  {author} {\bibinfo {author} {\bibfnamefont {R.~L.}\ \bibnamefont {Workman}}\ and\ \bibinfo {author} {\bibnamefont {Others}} (\bibinfo {collaboration} {Particle Data Group}),\ }\bibfield  {title} {\bibinfo {title} {{Review of Particle Physics}},\ }\href {https://doi.org/10.1093/ptep/ptac097} {\bibfield  {journal} {\bibinfo  {journal} {PTEP}\ }\textbf {\bibinfo {volume} {2022}},\ \bibinfo {pages} {083C01} (\bibinfo {year} {2022})}\BibitemShut {NoStop}%
\bibitem [{\citenamefont {Aaij}\ \emph {et~al.}(2022)\citenamefont {Aaij} \emph {et~al.}}]{LHCb:2021bjt}%
  \BibitemOpen
  \bibfield  {author} {\bibinfo {author} {\bibfnamefont {R.}~\bibnamefont {Aaij}} \emph {et~al.} (\bibinfo {collaboration} {LHCb}),\ }\bibfield  {title} {\bibinfo {title} {{Measurement of the W boson mass}},\ }\href {https://doi.org/10.1007/JHEP01(2022)036} {\bibfield  {journal} {\bibinfo  {journal} {JHEP}\ }\textbf {\bibinfo {volume} {01}},\ \bibinfo {pages} {036}},\ \Eprint {https://arxiv.org/abs/2109.01113} {arXiv:2109.01113 [hep-ex]} \BibitemShut {NoStop}%
\bibitem [{\citenamefont {Aaltonen}\ \emph {et~al.}(2022)\citenamefont {Aaltonen} \emph {et~al.}}]{CDF:2022hxs}%
  \BibitemOpen
  \bibfield  {author} {\bibinfo {author} {\bibfnamefont {T.}~\bibnamefont {Aaltonen}} \emph {et~al.} (\bibinfo {collaboration} {CDF}),\ }\bibfield  {title} {\bibinfo {title} {{High-precision measurement of the $W$ boson mass with the CDF II detector}},\ }\href {https://doi.org/10.1126/science.abk1781} {\bibfield  {journal} {\bibinfo  {journal} {Science}\ }\textbf {\bibinfo {volume} {376}},\ \bibinfo {pages} {170} (\bibinfo {year} {2022})}\BibitemShut {NoStop}%
\bibitem [{\citenamefont {Aaboud}\ \emph {et~al.}(2018)\citenamefont {Aaboud} \emph {et~al.}}]{ATLAS:2017rzl}%
  \BibitemOpen
  \bibfield  {author} {\bibinfo {author} {\bibfnamefont {M.}~\bibnamefont {Aaboud}} \emph {et~al.} (\bibinfo {collaboration} {ATLAS}),\ }\bibfield  {title} {\bibinfo {title} {{Measurement of the $W$-boson mass in pp collisions at $\sqrt{s}=7$ TeV with the ATLAS detector}},\ }\href {https://doi.org/10.1140/epjc/s10052-017-5475-4} {\bibfield  {journal} {\bibinfo  {journal} {Eur. Phys. J. C}\ }\textbf {\bibinfo {volume} {78}},\ \bibinfo {pages} {110} (\bibinfo {year} {2018})},\ \bibinfo {note} {[Erratum: Eur.Phys.J.C 78, 898 (2018)]},\ \Eprint {https://arxiv.org/abs/1701.07240} {arXiv:1701.07240 [hep-ex]} \BibitemShut {NoStop}%
\bibitem [{\citenamefont {Aaltonen}\ \emph {et~al.}(2012)\citenamefont {Aaltonen} \emph {et~al.}}]{CDF:2012gpf}%
  \BibitemOpen
  \bibfield  {author} {\bibinfo {author} {\bibfnamefont {T.}~\bibnamefont {Aaltonen}} \emph {et~al.} (\bibinfo {collaboration} {CDF}),\ }\bibfield  {title} {\bibinfo {title} {{Precise measurement of the $W$-boson mass with the CDF II detector}},\ }\href {https://doi.org/10.1103/PhysRevLett.108.151803} {\bibfield  {journal} {\bibinfo  {journal} {Phys. Rev. Lett.}\ }\textbf {\bibinfo {volume} {108}},\ \bibinfo {pages} {151803} (\bibinfo {year} {2012})},\ \Eprint {https://arxiv.org/abs/1203.0275} {arXiv:1203.0275 [hep-ex]} \BibitemShut {NoStop}%
\bibitem [{\citenamefont {Abazov}\ \emph {et~al.}(2012)\citenamefont {Abazov} \emph {et~al.}}]{D0:2012kms}%
  \BibitemOpen
  \bibfield  {author} {\bibinfo {author} {\bibfnamefont {V.~M.}\ \bibnamefont {Abazov}} \emph {et~al.} (\bibinfo {collaboration} {D0}),\ }\bibfield  {title} {\bibinfo {title} {{Measurement of the W Boson Mass with the D0 Detector}},\ }\href {https://doi.org/10.1103/PhysRevLett.108.151804} {\bibfield  {journal} {\bibinfo  {journal} {Phys. Rev. Lett.}\ }\textbf {\bibinfo {volume} {108}},\ \bibinfo {pages} {151804} (\bibinfo {year} {2012})},\ \Eprint {https://arxiv.org/abs/1203.0293} {arXiv:1203.0293 [hep-ex]} \BibitemShut {NoStop}%
\bibitem [{\citenamefont {Abdallah}\ \emph {et~al.}(2008)\citenamefont {Abdallah} \emph {et~al.}}]{DELPHI:2008avl}%
  \BibitemOpen
  \bibfield  {author} {\bibinfo {author} {\bibfnamefont {J.}~\bibnamefont {Abdallah}} \emph {et~al.} (\bibinfo {collaboration} {DELPHI}),\ }\bibfield  {title} {\bibinfo {title} {{Measurement of the Mass and Width of the $W$ Boson in $e^{+} e^{-}$ Collisions at $\sqrt{s}$ = 161-GeV - 209-GeV}},\ }\href {https://doi.org/10.1140/epjc/s10052-008-0585-7} {\bibfield  {journal} {\bibinfo  {journal} {Eur. Phys. J. C}\ }\textbf {\bibinfo {volume} {55}},\ \bibinfo {pages} {1} (\bibinfo {year} {2008})},\ \Eprint {https://arxiv.org/abs/0803.2534} {arXiv:0803.2534 [hep-ex]} \BibitemShut {NoStop}%
\bibitem [{\citenamefont {Abbiendi}\ \emph {et~al.}(2006)\citenamefont {Abbiendi} \emph {et~al.}}]{OPAL:2005rdt}%
  \BibitemOpen
  \bibfield  {author} {\bibinfo {author} {\bibfnamefont {G.}~\bibnamefont {Abbiendi}} \emph {et~al.} (\bibinfo {collaboration} {OPAL}),\ }\bibfield  {title} {\bibinfo {title} {{Measurement of the mass and width of the $W$ boson}},\ }\href {https://doi.org/10.1140/epjc/s2005-02440-5} {\bibfield  {journal} {\bibinfo  {journal} {Eur. Phys. J. C}\ }\textbf {\bibinfo {volume} {45}},\ \bibinfo {pages} {307} (\bibinfo {year} {2006})},\ \Eprint {https://arxiv.org/abs/hep-ex/0508060} {arXiv:hep-ex/0508060} \BibitemShut {NoStop}%
\bibitem [{\citenamefont {Achard}\ \emph {et~al.}(2006)\citenamefont {Achard} \emph {et~al.}}]{L3:2005fft}%
  \BibitemOpen
  \bibfield  {author} {\bibinfo {author} {\bibfnamefont {P.}~\bibnamefont {Achard}} \emph {et~al.} (\bibinfo {collaboration} {L3}),\ }\bibfield  {title} {\bibinfo {title} {{Measurement of the mass and the width of the $W$ boson at LEP}},\ }\href {https://doi.org/10.1140/epjc/s2005-02459-6} {\bibfield  {journal} {\bibinfo  {journal} {Eur. Phys. J. C}\ }\textbf {\bibinfo {volume} {45}},\ \bibinfo {pages} {569} (\bibinfo {year} {2006})},\ \Eprint {https://arxiv.org/abs/hep-ex/0511049} {arXiv:hep-ex/0511049} \BibitemShut {NoStop}%
\bibitem [{\citenamefont {Schael}\ \emph {et~al.}(2006)\citenamefont {Schael} \emph {et~al.}}]{ALEPH:2006cdc}%
  \BibitemOpen
  \bibfield  {author} {\bibinfo {author} {\bibfnamefont {S.}~\bibnamefont {Schael}} \emph {et~al.} (\bibinfo {collaboration} {ALEPH}),\ }\bibfield  {title} {\bibinfo {title} {{Measurement of the $W$ boson mass and width in $e^{+} e^{-}$ collisions at LEP}},\ }\href {https://doi.org/10.1140/epjc/s2006-02576-8} {\bibfield  {journal} {\bibinfo  {journal} {Eur. Phys. J. C}\ }\textbf {\bibinfo {volume} {47}},\ \bibinfo {pages} {309} (\bibinfo {year} {2006})},\ \Eprint {https://arxiv.org/abs/hep-ex/0605011} {arXiv:hep-ex/0605011} \BibitemShut {NoStop}%
\bibitem [{\citenamefont {Abazov}\ \emph {et~al.}(2002)\citenamefont {Abazov} \emph {et~al.}}]{D0:2002fhu}%
  \BibitemOpen
  \bibfield  {author} {\bibinfo {author} {\bibfnamefont {V.~M.}\ \bibnamefont {Abazov}} \emph {et~al.} (\bibinfo {collaboration} {D0}),\ }\bibfield  {title} {\bibinfo {title} {{Improved $W$ Boson Mass Measurement with the D0 Detector}},\ }\href {https://doi.org/10.1103/PhysRevD.66.012001} {\bibfield  {journal} {\bibinfo  {journal} {Phys. Rev. D}\ }\textbf {\bibinfo {volume} {66}},\ \bibinfo {pages} {012001} (\bibinfo {year} {2002})},\ \Eprint {https://arxiv.org/abs/hep-ex/0204014} {arXiv:hep-ex/0204014} \BibitemShut {NoStop}%
\bibitem [{\citenamefont {Affolder}\ \emph {et~al.}(2001)\citenamefont {Affolder} \emph {et~al.}}]{CDF:2000gwd}%
  \BibitemOpen
  \bibfield  {author} {\bibinfo {author} {\bibfnamefont {T.}~\bibnamefont {Affolder}} \emph {et~al.} (\bibinfo {collaboration} {CDF}),\ }\bibfield  {title} {\bibinfo {title} {{Measurement of the $W$ boson mass with the Collider Detector at Fermilab}},\ }\href {https://doi.org/10.1103/PhysRevD.64.052001} {\bibfield  {journal} {\bibinfo  {journal} {Phys. Rev. D}\ }\textbf {\bibinfo {volume} {64}},\ \bibinfo {pages} {052001} (\bibinfo {year} {2001})},\ \Eprint {https://arxiv.org/abs/hep-ex/0007044} {arXiv:hep-ex/0007044} \BibitemShut {NoStop}%
\bibitem [{\citenamefont {Tumasyan}\ \emph {et~al.}(2022{\natexlab{a}})\citenamefont {Tumasyan} \emph {et~al.}}]{CMS:2022mhs}%
  \BibitemOpen
  \bibfield  {author} {\bibinfo {author} {\bibfnamefont {A.}~\bibnamefont {Tumasyan}} \emph {et~al.} (\bibinfo {collaboration} {CMS}),\ }\bibfield  {title} {\bibinfo {title} {{Precision measurement of the W boson decay branching fractions in proton-proton collisions at $\sqrt{s}$ = 13 TeV}},\ }\href {https://doi.org/10.1103/PhysRevD.105.072008} {\bibfield  {journal} {\bibinfo  {journal} {Phys. Rev. D}\ }\textbf {\bibinfo {volume} {105}},\ \bibinfo {pages} {072008} (\bibinfo {year} {2022}{\natexlab{a}})},\ \Eprint {https://arxiv.org/abs/2201.07861} {arXiv:2201.07861 [hep-ex]} \BibitemShut {NoStop}%
\bibitem [{\citenamefont {Abbiendi}\ \emph {et~al.}(2007)\citenamefont {Abbiendi} \emph {et~al.}}]{OPAL:2007ytu}%
  \BibitemOpen
  \bibfield  {author} {\bibinfo {author} {\bibfnamefont {G.}~\bibnamefont {Abbiendi}} \emph {et~al.} (\bibinfo {collaboration} {OPAL}),\ }\bibfield  {title} {\bibinfo {title} {{Measurement of the e+ e- ---\ensuremath{>} W+ W- cross section and W decay branching fractions at LEP}},\ }\href {https://doi.org/10.1140/epjc/s10052-007-0442-0} {\bibfield  {journal} {\bibinfo  {journal} {Eur. Phys. J. C}\ }\textbf {\bibinfo {volume} {52}},\ \bibinfo {pages} {767} (\bibinfo {year} {2007})},\ \Eprint {https://arxiv.org/abs/0708.1311} {arXiv:0708.1311 [hep-ex]} \BibitemShut {NoStop}%
\bibitem [{\citenamefont {Abdallah}\ \emph {et~al.}(2004)\citenamefont {Abdallah} \emph {et~al.}}]{DELPHI:2003ftu}%
  \BibitemOpen
  \bibfield  {author} {\bibinfo {author} {\bibfnamefont {J.}~\bibnamefont {Abdallah}} \emph {et~al.} (\bibinfo {collaboration} {DELPHI}),\ }\bibfield  {title} {\bibinfo {title} {{Measurement of the W pair production cross-section and W branching ratios in e+ e- collisions at s**(1/2) = 161-GeV to 209-GeV}},\ }\href {https://doi.org/10.1140/epjc/s2004-01709-5} {\bibfield  {journal} {\bibinfo  {journal} {Eur. Phys. J. C}\ }\textbf {\bibinfo {volume} {34}},\ \bibinfo {pages} {127} (\bibinfo {year} {2004})},\ \Eprint {https://arxiv.org/abs/hep-ex/0403042} {arXiv:hep-ex/0403042} \BibitemShut {NoStop}%
\bibitem [{\citenamefont {Achard}\ \emph {et~al.}(2004)\citenamefont {Achard} \emph {et~al.}}]{L3:2004lwm}%
  \BibitemOpen
  \bibfield  {author} {\bibinfo {author} {\bibfnamefont {P.}~\bibnamefont {Achard}} \emph {et~al.} (\bibinfo {collaboration} {L3}),\ }\bibfield  {title} {\bibinfo {title} {{Measurement of the cross section of W-boson pair production at LEP}},\ }\href {https://doi.org/10.1016/j.physletb.2004.08.060} {\bibfield  {journal} {\bibinfo  {journal} {Phys. Lett. B}\ }\textbf {\bibinfo {volume} {600}},\ \bibinfo {pages} {22} (\bibinfo {year} {2004})},\ \Eprint {https://arxiv.org/abs/hep-ex/0409016} {arXiv:hep-ex/0409016} \BibitemShut {NoStop}%
\bibitem [{\citenamefont {Heister}\ \emph {et~al.}(2004)\citenamefont {Heister} \emph {et~al.}}]{ALEPH:2004dmh}%
  \BibitemOpen
  \bibfield  {author} {\bibinfo {author} {\bibfnamefont {A.}~\bibnamefont {Heister}} \emph {et~al.} (\bibinfo {collaboration} {ALEPH}),\ }\bibfield  {title} {\bibinfo {title} {{Measurement of W-pair production in e+ e- collisions at centre-of-mass energies from 183-GeV to 209-GeV}},\ }\href {https://doi.org/10.1140/epjc/s2004-02048-3} {\bibfield  {journal} {\bibinfo  {journal} {Eur. Phys. J. C}\ }\textbf {\bibinfo {volume} {38}},\ \bibinfo {pages} {147} (\bibinfo {year} {2004})}\BibitemShut {NoStop}%
\bibitem [{\citenamefont {Abbiendi}\ \emph {et~al.}(2001{\natexlab{a}})\citenamefont {Abbiendi} \emph {et~al.}}]{OPAL:2000ufp}%
  \BibitemOpen
  \bibfield  {author} {\bibinfo {author} {\bibfnamefont {G.}~\bibnamefont {Abbiendi}} \emph {et~al.} (\bibinfo {collaboration} {OPAL}),\ }\bibfield  {title} {\bibinfo {title} {{Precise determination of the Z resonance parameters at LEP: 'Zedometry'}},\ }\href {https://doi.org/10.1007/s100520100627} {\bibfield  {journal} {\bibinfo  {journal} {Eur. Phys. J. C}\ }\textbf {\bibinfo {volume} {19}},\ \bibinfo {pages} {587} (\bibinfo {year} {2001}{\natexlab{a}})},\ \Eprint {https://arxiv.org/abs/hep-ex/0012018} {arXiv:hep-ex/0012018} \BibitemShut {NoStop}%
\bibitem [{\citenamefont {Abreu}\ \emph {et~al.}(2000{\natexlab{a}})\citenamefont {Abreu} \emph {et~al.}}]{DELPHI:2000wje}%
  \BibitemOpen
  \bibfield  {author} {\bibinfo {author} {\bibfnamefont {P.}~\bibnamefont {Abreu}} \emph {et~al.} (\bibinfo {collaboration} {DELPHI}),\ }\bibfield  {title} {\bibinfo {title} {{Cross-sections and leptonic forward backward asymmetries from the Z0 running of LEP}},\ }\href {https://doi.org/10.1007/s100520000392} {\bibfield  {journal} {\bibinfo  {journal} {Eur. Phys. J. C}\ }\textbf {\bibinfo {volume} {16}},\ \bibinfo {pages} {371} (\bibinfo {year} {2000}{\natexlab{a}})}\BibitemShut {NoStop}%
\bibitem [{\citenamefont {Acciarri}\ \emph {et~al.}(2000)\citenamefont {Acciarri} \emph {et~al.}}]{L3:2000vgx}%
  \BibitemOpen
  \bibfield  {author} {\bibinfo {author} {\bibfnamefont {M.}~\bibnamefont {Acciarri}} \emph {et~al.} (\bibinfo {collaboration} {L3}),\ }\bibfield  {title} {\bibinfo {title} {{Measurements of cross-sections and forward backward asymmetries at the $Z$ resonance and determination of electroweak parameters}},\ }\href {https://doi.org/10.1007/s100520050001} {\bibfield  {journal} {\bibinfo  {journal} {Eur. Phys. J. C}\ }\textbf {\bibinfo {volume} {16}},\ \bibinfo {pages} {1} (\bibinfo {year} {2000})},\ \Eprint {https://arxiv.org/abs/hep-ex/0002046} {arXiv:hep-ex/0002046} \BibitemShut {NoStop}%
\bibitem [{\citenamefont {Barate}\ \emph {et~al.}(2000)\citenamefont {Barate} \emph {et~al.}}]{ALEPH:1999smx}%
  \BibitemOpen
  \bibfield  {author} {\bibinfo {author} {\bibfnamefont {R.}~\bibnamefont {Barate}} \emph {et~al.} (\bibinfo {collaboration} {ALEPH}),\ }\bibfield  {title} {\bibinfo {title} {{Measurement of the Z resonance parameters at LEP}},\ }\href {https://doi.org/10.1007/s100520000319} {\bibfield  {journal} {\bibinfo  {journal} {Eur. Phys. J. C}\ }\textbf {\bibinfo {volume} {14}},\ \bibinfo {pages} {1} (\bibinfo {year} {2000})}\BibitemShut {NoStop}%
\bibitem [{\citenamefont {Abbiendi}\ \emph {et~al.}(2001{\natexlab{b}})\citenamefont {Abbiendi} \emph {et~al.}}]{OPAL:2001brm}%
  \BibitemOpen
  \bibfield  {author} {\bibinfo {author} {\bibfnamefont {G.}~\bibnamefont {Abbiendi}} \emph {et~al.} (\bibinfo {collaboration} {OPAL}),\ }\bibfield  {title} {\bibinfo {title} {{Precision neutral current asymmetry parameter measurements from the tau polarization at LEP}},\ }\href {https://doi.org/10.1007/s100520100714} {\bibfield  {journal} {\bibinfo  {journal} {Eur. Phys. J. C}\ }\textbf {\bibinfo {volume} {21}},\ \bibinfo {pages} {1} (\bibinfo {year} {2001}{\natexlab{b}})},\ \Eprint {https://arxiv.org/abs/hep-ex/0103045} {arXiv:hep-ex/0103045} \BibitemShut {NoStop}%
\bibitem [{\citenamefont {Abe}\ \emph {et~al.}(2001)\citenamefont {Abe} \emph {et~al.}}]{SLD:2000ujp}%
  \BibitemOpen
  \bibfield  {author} {\bibinfo {author} {\bibfnamefont {K.}~\bibnamefont {Abe}} \emph {et~al.} (\bibinfo {collaboration} {SLD}),\ }\bibfield  {title} {\bibinfo {title} {{An Improved direct measurement of leptonic coupling asymmetries with polarized Z bosons}},\ }\href {https://doi.org/10.1103/PhysRevLett.86.1162} {\bibfield  {journal} {\bibinfo  {journal} {Phys. Rev. Lett.}\ }\textbf {\bibinfo {volume} {86}},\ \bibinfo {pages} {1162} (\bibinfo {year} {2001})},\ \Eprint {https://arxiv.org/abs/hep-ex/0010015} {arXiv:hep-ex/0010015} \BibitemShut {NoStop}%
\bibitem [{\citenamefont {Heister}\ \emph {et~al.}(2001)\citenamefont {Heister} \emph {et~al.}}]{ALEPH:2001uca}%
  \BibitemOpen
  \bibfield  {author} {\bibinfo {author} {\bibfnamefont {A.}~\bibnamefont {Heister}} \emph {et~al.} (\bibinfo {collaboration} {ALEPH}),\ }\bibfield  {title} {\bibinfo {title} {{Measurement of the tau polarization at LEP}},\ }\href {https://doi.org/10.1007/s100520100689} {\bibfield  {journal} {\bibinfo  {journal} {Eur. Phys. J. C}\ }\textbf {\bibinfo {volume} {20}},\ \bibinfo {pages} {401} (\bibinfo {year} {2001})},\ \Eprint {https://arxiv.org/abs/hep-ex/0104038} {arXiv:hep-ex/0104038} \BibitemShut {NoStop}%
\bibitem [{\citenamefont {Abreu}\ \emph {et~al.}(2000{\natexlab{b}})\citenamefont {Abreu} \emph {et~al.}}]{DELPHI:1999yne}%
  \BibitemOpen
  \bibfield  {author} {\bibinfo {author} {\bibfnamefont {P.}~\bibnamefont {Abreu}} \emph {et~al.} (\bibinfo {collaboration} {DELPHI}),\ }\bibfield  {title} {\bibinfo {title} {{A Precise measurement of the tau polarization at LEP-1}},\ }\href {https://doi.org/10.1007/s100520000363} {\bibfield  {journal} {\bibinfo  {journal} {Eur. Phys. J. C}\ }\textbf {\bibinfo {volume} {14}},\ \bibinfo {pages} {585} (\bibinfo {year} {2000}{\natexlab{b}})}\BibitemShut {NoStop}%
\bibitem [{\citenamefont {Acciarri}\ \emph {et~al.}(1998)\citenamefont {Acciarri} \emph {et~al.}}]{L3:1998oan}%
  \BibitemOpen
  \bibfield  {author} {\bibinfo {author} {\bibfnamefont {M.}~\bibnamefont {Acciarri}} \emph {et~al.} (\bibinfo {collaboration} {L3}),\ }\bibfield  {title} {\bibinfo {title} {{Measurement of tau polarization at LEP}},\ }\href {https://doi.org/10.1016/S0370-2693(98)00406-7} {\bibfield  {journal} {\bibinfo  {journal} {Phys. Lett. B}\ }\textbf {\bibinfo {volume} {429}},\ \bibinfo {pages} {387} (\bibinfo {year} {1998})}\BibitemShut {NoStop}%
\bibitem [{\citenamefont {Abe}\ \emph {et~al.}(1997)\citenamefont {Abe} \emph {et~al.}}]{SLD:1996gjt}%
  \BibitemOpen
  \bibfield  {author} {\bibinfo {author} {\bibfnamefont {K.}~\bibnamefont {Abe}} \emph {et~al.} (\bibinfo {collaboration} {SLD}),\ }\bibfield  {title} {\bibinfo {title} {{First measurement of the left-right charge asymmetry in hadronic Z boson decays and a new determination of sin**2 theta(W)(eff)}},\ }\href {https://doi.org/10.1103/PhysRevLett.78.17} {\bibfield  {journal} {\bibinfo  {journal} {Phys. Rev. Lett.}\ }\textbf {\bibinfo {volume} {78}},\ \bibinfo {pages} {17} (\bibinfo {year} {1997})},\ \Eprint {https://arxiv.org/abs/hep-ex/9609019} {arXiv:hep-ex/9609019} \BibitemShut {NoStop}%
\bibitem [{\citenamefont {Abe}\ \emph {et~al.}(1995)\citenamefont {Abe} \emph {et~al.}}]{SLD:1994kvj}%
  \BibitemOpen
  \bibfield  {author} {\bibinfo {author} {\bibfnamefont {K.}~\bibnamefont {Abe}} \emph {et~al.} (\bibinfo {collaboration} {SLD}),\ }\bibfield  {title} {\bibinfo {title} {{Polarized Bhabha scattering a precision measurement of the electron neutral current couplings}},\ }\href {https://doi.org/10.1103/PhysRevLett.74.2880} {\bibfield  {journal} {\bibinfo  {journal} {Phys. Rev. Lett.}\ }\textbf {\bibinfo {volume} {74}},\ \bibinfo {pages} {2880} (\bibinfo {year} {1995})},\ \Eprint {https://arxiv.org/abs/hep-ex/9410009} {arXiv:hep-ex/9410009} \BibitemShut {NoStop}%
\bibitem [{\citenamefont {{Parker}}\ \emph {et~al.}(2018)\citenamefont {{Parker}}, \citenamefont {{Yu}}, \citenamefont {{Zhong}}, \citenamefont {{Estey}},\ and\ \citenamefont {{M{\"u}ller}}}]{2018Sci...360..191P}%
  \BibitemOpen
  \bibfield  {author} {\bibinfo {author} {\bibfnamefont {R.~H.}\ \bibnamefont {{Parker}}}, \bibinfo {author} {\bibfnamefont {C.}~\bibnamefont {{Yu}}}, \bibinfo {author} {\bibfnamefont {W.}~\bibnamefont {{Zhong}}}, \bibinfo {author} {\bibfnamefont {B.}~\bibnamefont {{Estey}}},\ and\ \bibinfo {author} {\bibfnamefont {H.}~\bibnamefont {{M{\"u}ller}}},\ }\bibfield  {title} {\bibinfo {title} {{Measurement of the fine-structure constant as a test of the Standard Model}},\ }\href {https://doi.org/10.1126/science.aap7706} {\bibfield  {journal} {\bibinfo  {journal} {Science}\ }\textbf {\bibinfo {volume} {360}},\ \bibinfo {pages} {191} (\bibinfo {year} {2018})},\ \Eprint {https://arxiv.org/abs/1812.04130} {arXiv:1812.04130 [physics.atom-ph]} \BibitemShut {NoStop}%
\bibitem [{\citenamefont {Aad}\ \emph {et~al.}(2022)\citenamefont {Aad} \emph {et~al.}}]{ATLAS:2022vkf}%
  \BibitemOpen
  \bibfield  {author} {\bibinfo {author} {\bibfnamefont {G.}~\bibnamefont {Aad}} \emph {et~al.} (\bibinfo {collaboration} {ATLAS}),\ }\bibfield  {title} {\bibinfo {title} {{A detailed map of Higgs boson interactions by the ATLAS experiment ten years after the discovery}},\ }\href {https://doi.org/10.1038/s41586-022-04893-w} {\bibfield  {journal} {\bibinfo  {journal} {Nature}\ }\textbf {\bibinfo {volume} {607}},\ \bibinfo {pages} {52} (\bibinfo {year} {2022})},\ \bibinfo {note} {[Erratum: Nature 612, E24 (2022)]},\ \Eprint {https://arxiv.org/abs/2207.00092} {arXiv:2207.00092 [hep-ex]} \BibitemShut {NoStop}%
\bibitem [{\citenamefont {Tumasyan}\ \emph {et~al.}(2022{\natexlab{b}})\citenamefont {Tumasyan} \emph {et~al.}}]{CMS:2022dwd}%
  \BibitemOpen
  \bibfield  {author} {\bibinfo {author} {\bibfnamefont {A.}~\bibnamefont {Tumasyan}} \emph {et~al.} (\bibinfo {collaboration} {CMS}),\ }\bibfield  {title} {\bibinfo {title} {{A portrait of the Higgs boson by the CMS experiment ten years after the discovery.}},\ }\href {https://doi.org/10.1038/s41586-022-04892-x} {\bibfield  {journal} {\bibinfo  {journal} {Nature}\ }\textbf {\bibinfo {volume} {607}},\ \bibinfo {pages} {60} (\bibinfo {year} {2022}{\natexlab{b}})},\ \bibinfo {note} {[Erratum: Nature 623, (2023)]},\ \Eprint {https://arxiv.org/abs/2207.00043} {arXiv:2207.00043 [hep-ex]} \BibitemShut {NoStop}%
\bibitem [{\citenamefont {Baldini}\ \emph {et~al.}(2016)\citenamefont {Baldini} \emph {et~al.}}]{MEG:2016leq}%
  \BibitemOpen
  \bibfield  {author} {\bibinfo {author} {\bibfnamefont {A.~M.}\ \bibnamefont {Baldini}} \emph {et~al.} (\bibinfo {collaboration} {MEG}),\ }\bibfield  {title} {\bibinfo {title} {{Search for the lepton flavour violating decay $\mu ^+ \rightarrow \mathrm {e}^+ \gamma $ with the full dataset of the MEG experiment}},\ }\href {https://doi.org/10.1140/epjc/s10052-016-4271-x} {\bibfield  {journal} {\bibinfo  {journal} {Eur. Phys. J. C}\ }\textbf {\bibinfo {volume} {76}},\ \bibinfo {pages} {434} (\bibinfo {year} {2016})},\ \Eprint {https://arxiv.org/abs/1605.05081} {arXiv:1605.05081 [hep-ex]} \BibitemShut {NoStop}%
\bibitem [{\citenamefont {Aubert}\ \emph {et~al.}(2010)\citenamefont {Aubert} \emph {et~al.}}]{BaBar:2009hkt}%
  \BibitemOpen
  \bibfield  {author} {\bibinfo {author} {\bibfnamefont {B.}~\bibnamefont {Aubert}} \emph {et~al.} (\bibinfo {collaboration} {BaBar}),\ }\bibfield  {title} {\bibinfo {title} {{Searches for Lepton Flavor Violation in the Decays tau+- ---\ensuremath{>} e+- gamma and tau+- ---\ensuremath{>} mu+- gamma}},\ }\href {https://doi.org/10.1103/PhysRevLett.104.021802} {\bibfield  {journal} {\bibinfo  {journal} {Phys. Rev. Lett.}\ }\textbf {\bibinfo {volume} {104}},\ \bibinfo {pages} {021802} (\bibinfo {year} {2010})},\ \Eprint {https://arxiv.org/abs/0908.2381} {arXiv:0908.2381 [hep-ex]} \BibitemShut {NoStop}%
\bibitem [{\citenamefont {Abdesselam}\ \emph {et~al.}(2021)\citenamefont {Abdesselam} \emph {et~al.}}]{Belle:2021ysv}%
  \BibitemOpen
  \bibfield  {author} {\bibinfo {author} {\bibfnamefont {A.}~\bibnamefont {Abdesselam}} \emph {et~al.} (\bibinfo {collaboration} {Belle}),\ }\bibfield  {title} {\bibinfo {title} {{Search for lepton-flavor-violating tau-lepton decays to $\ell\gamma$ at Belle}},\ }\href {https://doi.org/10.1007/JHEP10(2021)019} {\bibfield  {journal} {\bibinfo  {journal} {JHEP}\ }\textbf {\bibinfo {volume} {10}},\ \bibinfo {pages} {19}},\ \Eprint {https://arxiv.org/abs/2103.12994} {arXiv:2103.12994 [hep-ex]} \BibitemShut {NoStop}%
\bibitem [{\citenamefont {Bertl}\ \emph {et~al.}(2006)\citenamefont {Bertl} \emph {et~al.}}]{SINDRUMII:2006dvw}%
  \BibitemOpen
  \bibfield  {author} {\bibinfo {author} {\bibfnamefont {W.~H.}\ \bibnamefont {Bertl}} \emph {et~al.} (\bibinfo {collaboration} {SINDRUM II}),\ }\bibfield  {title} {\bibinfo {title} {{A Search for muon to electron conversion in muonic gold}},\ }\href {https://doi.org/10.1140/epjc/s2006-02582-x} {\bibfield  {journal} {\bibinfo  {journal} {Eur. Phys. J. C}\ }\textbf {\bibinfo {volume} {47}},\ \bibinfo {pages} {337} (\bibinfo {year} {2006})}\BibitemShut {NoStop}%
\bibitem [{\citenamefont {{Fan}}\ \emph {et~al.}(2019)\citenamefont {{Fan}}, \citenamefont {{Su}}, \citenamefont {{Zhang}},\ and\ \citenamefont {{Yu}}}]{2019arXiv190301344F}%
  \BibitemOpen
  \bibfield  {author} {\bibinfo {author} {\bibfnamefont {Z.}~\bibnamefont {{Fan}}}, \bibinfo {author} {\bibfnamefont {R.}~\bibnamefont {{Su}}}, \bibinfo {author} {\bibfnamefont {W.}~\bibnamefont {{Zhang}}},\ and\ \bibinfo {author} {\bibfnamefont {Y.}~\bibnamefont {{Yu}}},\ }\bibfield  {title} {\bibinfo {title} {{Hybrid Actor-Critic Reinforcement Learning in Parameterized Action Space}},\ }\href {https://doi.org/10.48550/arXiv.1903.01344} {\bibfield  {journal} {\bibinfo  {journal} {arXiv e-prints}\ ,\ \bibinfo {eid} {arXiv:1903.01344}} (\bibinfo {year} {2019})},\ \Eprint {https://arxiv.org/abs/1903.01344} {arXiv:1903.01344 [cs.LG]} \BibitemShut {NoStop}%
\bibitem [{\citenamefont {Towers}\ \emph {et~al.}(2023)\citenamefont {Towers}, \citenamefont {Terry}, \citenamefont {Kwiatkowski}, \citenamefont {Balis}, \citenamefont {Cola}, \citenamefont {Deleu}, \citenamefont {Goulão}, \citenamefont {Kallinteris}, \citenamefont {KG}, \citenamefont {Krimmel}, \citenamefont {Perez-Vicente}, \citenamefont {Pierré}, \citenamefont {Schulhoff}, \citenamefont {Tai}, \citenamefont {Shen},\ and\ \citenamefont {Younis}}]{towers_gymnasium_2023}%
  \BibitemOpen
  \bibfield  {author} {\bibinfo {author} {\bibfnamefont {M.}~\bibnamefont {Towers}}, \bibinfo {author} {\bibfnamefont {J.~K.}\ \bibnamefont {Terry}}, \bibinfo {author} {\bibfnamefont {A.}~\bibnamefont {Kwiatkowski}}, \bibinfo {author} {\bibfnamefont {J.~U.}\ \bibnamefont {Balis}}, \bibinfo {author} {\bibfnamefont {G.~d.}\ \bibnamefont {Cola}}, \bibinfo {author} {\bibfnamefont {T.}~\bibnamefont {Deleu}}, \bibinfo {author} {\bibfnamefont {M.}~\bibnamefont {Goulão}}, \bibinfo {author} {\bibfnamefont {A.}~\bibnamefont {Kallinteris}}, \bibinfo {author} {\bibfnamefont {A.}~\bibnamefont {KG}}, \bibinfo {author} {\bibfnamefont {M.}~\bibnamefont {Krimmel}}, \bibinfo {author} {\bibfnamefont {R.}~\bibnamefont {Perez-Vicente}}, \bibinfo {author} {\bibfnamefont {A.}~\bibnamefont {Pierré}}, \bibinfo {author} {\bibfnamefont {S.}~\bibnamefont {Schulhoff}}, \bibinfo {author} {\bibfnamefont {J.~J.}\ \bibnamefont {Tai}}, \bibinfo {author} {\bibfnamefont {A.~T.~J.}\ \bibnamefont {Shen}},\ and\ \bibinfo {author} {\bibfnamefont
  {O.~G.}\ \bibnamefont {Younis}},\ }\href {https://doi.org/10.5281/zenodo.8127026} {\bibinfo {title} {Gymnasium}} (\bibinfo {year} {2023})\BibitemShut {NoStop}%
\bibitem [{\citenamefont {{Fey}}\ and\ \citenamefont {{Lenssen}}(2019)}]{2019arXiv190302428F}%
  \BibitemOpen
  \bibfield  {author} {\bibinfo {author} {\bibfnamefont {M.}~\bibnamefont {{Fey}}}\ and\ \bibinfo {author} {\bibfnamefont {J.~E.}\ \bibnamefont {{Lenssen}}},\ }\bibfield  {title} {\bibinfo {title} {{Fast Graph Representation Learning with PyTorch Geometric}},\ }\href {https://doi.org/10.48550/arXiv.1903.02428} {\bibfield  {journal} {\bibinfo  {journal} {arXiv e-prints}\ ,\ \bibinfo {eid} {arXiv:1903.02428}} (\bibinfo {year} {2019})},\ \Eprint {https://arxiv.org/abs/1903.02428} {arXiv:1903.02428 [cs.LG]} \BibitemShut {NoStop}%
\bibitem [{\citenamefont {Osman~Acar}\ \emph {et~al.}(2021)\citenamefont {Osman~Acar}, \citenamefont {Delialioglu},\ and\ \citenamefont {Sultansoy}}]{OsmanAcar:2021plv}%
  \BibitemOpen
  \bibfield  {author} {\bibinfo {author} {\bibfnamefont {A.}~\bibnamefont {Osman~Acar}}, \bibinfo {author} {\bibfnamefont {O.~E.}\ \bibnamefont {Delialioglu}},\ and\ \bibinfo {author} {\bibfnamefont {S.}~\bibnamefont {Sultansoy}},\ }\bibfield  {title} {\bibinfo {title} {{A search for the first generation charged vector-like leptons at future colliders}},\ }\href@noop {} {\  (\bibinfo {year} {2021})},\ \Eprint {https://arxiv.org/abs/2103.08222} {arXiv:2103.08222 [hep-ph]} \BibitemShut {NoStop}%
\bibitem [{\citenamefont {Aad}\ \emph {et~al.}(2015)\citenamefont {Aad} \emph {et~al.}}]{ATLAS:2015qoy}%
  \BibitemOpen
  \bibfield  {author} {\bibinfo {author} {\bibfnamefont {G.}~\bibnamefont {Aad}} \emph {et~al.} (\bibinfo {collaboration} {ATLAS}),\ }\bibfield  {title} {\bibinfo {title} {{Search for heavy lepton resonances decaying to a $Z$ boson and a lepton in $pp$ collisions at $\sqrt{s}=8$ TeV with the ATLAS detector}},\ }\href {https://doi.org/10.1007/JHEP09(2015)108} {\bibfield  {journal} {\bibinfo  {journal} {JHEP}\ }\textbf {\bibinfo {volume} {09}},\ \bibinfo {pages} {108}},\ \Eprint {https://arxiv.org/abs/1506.01291} {arXiv:1506.01291 [hep-ex]} \BibitemShut {NoStop}%
\bibitem [{\citenamefont {Aad}\ \emph {et~al.}(2023)\citenamefont {Aad} \emph {et~al.}}]{ATLAS:2023sbu}%
  \BibitemOpen
  \bibfield  {author} {\bibinfo {author} {\bibfnamefont {G.}~\bibnamefont {Aad}} \emph {et~al.} (\bibinfo {collaboration} {ATLAS}),\ }\bibfield  {title} {\bibinfo {title} {{Search for third-generation vector-like leptons in $pp$ collisions at $\sqrt{s} = 13\,\text{TeV}$ with the ATLAS detector}},\ }\href {https://doi.org/10.1007/JHEP07(2023)118} {\bibfield  {journal} {\bibinfo  {journal} {JHEP}\ }\textbf {\bibinfo {volume} {07}},\ \bibinfo {pages} {118}},\ \Eprint {https://arxiv.org/abs/2303.05441} {arXiv:2303.05441 [hep-ex]} \BibitemShut {NoStop}%
\bibitem [{\citenamefont {Guedes}\ and\ \citenamefont {Santiago}(2022)}]{Guedes:2021oqx}%
  \BibitemOpen
  \bibfield  {author} {\bibinfo {author} {\bibfnamefont {G.}~\bibnamefont {Guedes}}\ and\ \bibinfo {author} {\bibfnamefont {J.}~\bibnamefont {Santiago}},\ }\bibfield  {title} {\bibinfo {title} {{New leptons with exotic decays: collider limits and dark matter complementarity}},\ }\href {https://doi.org/10.1007/JHEP01(2022)111} {\bibfield  {journal} {\bibinfo  {journal} {JHEP}\ }\textbf {\bibinfo {volume} {01}},\ \bibinfo {pages} {111}},\ \Eprint {https://arxiv.org/abs/2107.03429} {arXiv:2107.03429 [hep-ph]} \BibitemShut {NoStop}%
\bibitem [{\citenamefont {Kawamura}\ and\ \citenamefont {Shin}(2023)}]{Kawamura:2023zuo}%
  \BibitemOpen
  \bibfield  {author} {\bibinfo {author} {\bibfnamefont {J.}~\bibnamefont {Kawamura}}\ and\ \bibinfo {author} {\bibfnamefont {S.}~\bibnamefont {Shin}},\ }\bibfield  {title} {\bibinfo {title} {{Current status on pair-produced muon-philic vectorlike leptons in multilepton channels at the LHC}},\ }\href {https://doi.org/10.1007/JHEP11(2023)025} {\bibfield  {journal} {\bibinfo  {journal} {JHEP}\ }\textbf {\bibinfo {volume} {11}},\ \bibinfo {pages} {025}},\ \Eprint {https://arxiv.org/abs/2308.07814} {arXiv:2308.07814 [hep-ph]} \BibitemShut {NoStop}%
\bibitem [{\citenamefont {Bhattiprolu}\ and\ \citenamefont {Martin}(2019)}]{Bhattiprolu:2019vdu}%
  \BibitemOpen
  \bibfield  {author} {\bibinfo {author} {\bibfnamefont {P.~N.}\ \bibnamefont {Bhattiprolu}}\ and\ \bibinfo {author} {\bibfnamefont {S.~P.}\ \bibnamefont {Martin}},\ }\bibfield  {title} {\bibinfo {title} {{Prospects for vectorlike leptons at future proton-proton colliders}},\ }\href {https://doi.org/10.1103/PhysRevD.100.015033} {\bibfield  {journal} {\bibinfo  {journal} {Phys. Rev. D}\ }\textbf {\bibinfo {volume} {100}},\ \bibinfo {pages} {015033} (\bibinfo {year} {2019})},\ \Eprint {https://arxiv.org/abs/1905.00498} {arXiv:1905.00498 [hep-ph]} \BibitemShut {NoStop}%
\bibitem [{\citenamefont {Straub}(2018)}]{Straub:2018kue}%
  \BibitemOpen
  \bibfield  {author} {\bibinfo {author} {\bibfnamefont {D.~M.}\ \bibnamefont {Straub}},\ }\bibfield  {title} {\bibinfo {title} {{flavio: a Python package for flavour and precision phenomenology in the Standard Model and beyond}},\ }\href@noop {} {\  (\bibinfo {year} {2018})},\ \Eprint {https://arxiv.org/abs/1810.08132} {arXiv:1810.08132 [hep-ph]} \BibitemShut {NoStop}%
\bibitem [{\citenamefont {Uhlrich}\ \emph {et~al.}(2021)\citenamefont {Uhlrich}, \citenamefont {Mahmoudi},\ and\ \citenamefont {Arbey}}]{Uhlrich:2020ltd}%
  \BibitemOpen
  \bibfield  {author} {\bibinfo {author} {\bibfnamefont {G.}~\bibnamefont {Uhlrich}}, \bibinfo {author} {\bibfnamefont {F.}~\bibnamefont {Mahmoudi}},\ and\ \bibinfo {author} {\bibfnamefont {A.}~\bibnamefont {Arbey}},\ }\bibfield  {title} {\bibinfo {title} {{MARTY -- Modern ARtificial Theoretical phYsicist A C++ framework automating symbolic calculations Beyond the Standard Model}},\ }\href {https://doi.org/10.1016/j.cpc.2021.107928} {\bibfield  {journal} {\bibinfo  {journal} {Comput. Phys. Commun.}\ }\textbf {\bibinfo {volume} {264}},\ \bibinfo {pages} {107928} (\bibinfo {year} {2021})},\ \Eprint {https://arxiv.org/abs/2011.02478} {arXiv:2011.02478 [hep-ph]} \BibitemShut {NoStop}%
\bibitem [{\citenamefont {Belanger}\ \emph {et~al.}(2010)\citenamefont {Belanger}, \citenamefont {Boudjema}, \citenamefont {Pukhov},\ and\ \citenamefont {Semenov}}]{Belanger:2010pz}%
  \BibitemOpen
  \bibfield  {author} {\bibinfo {author} {\bibfnamefont {G.}~\bibnamefont {Belanger}}, \bibinfo {author} {\bibfnamefont {F.}~\bibnamefont {Boudjema}}, \bibinfo {author} {\bibfnamefont {A.}~\bibnamefont {Pukhov}},\ and\ \bibinfo {author} {\bibfnamefont {A.}~\bibnamefont {Semenov}},\ }\bibfield  {title} {\bibinfo {title} {{micrOMEGAs: A Tool for dark matter studies}},\ }\href {https://doi.org/10.1393/ncc/i2010-10591-3} {\bibfield  {journal} {\bibinfo  {journal} {Nuovo Cim. C}\ }\textbf {\bibinfo {volume} {033N2}},\ \bibinfo {pages} {111} (\bibinfo {year} {2010})},\ \Eprint {https://arxiv.org/abs/1005.4133} {arXiv:1005.4133 [hep-ph]} \BibitemShut {NoStop}%
\bibitem [{\citenamefont {Alwall}\ \emph {et~al.}(2014)\citenamefont {Alwall}, \citenamefont {Frederix}, \citenamefont {Frixione}, \citenamefont {Hirschi}, \citenamefont {Maltoni}, \citenamefont {Mattelaer}, \citenamefont {Shao}, \citenamefont {Stelzer}, \citenamefont {Torrielli},\ and\ \citenamefont {Zaro}}]{Alwall:2014hca}%
  \BibitemOpen
  \bibfield  {author} {\bibinfo {author} {\bibfnamefont {J.}~\bibnamefont {Alwall}}, \bibinfo {author} {\bibfnamefont {R.}~\bibnamefont {Frederix}}, \bibinfo {author} {\bibfnamefont {S.}~\bibnamefont {Frixione}}, \bibinfo {author} {\bibfnamefont {V.}~\bibnamefont {Hirschi}}, \bibinfo {author} {\bibfnamefont {F.}~\bibnamefont {Maltoni}}, \bibinfo {author} {\bibfnamefont {O.}~\bibnamefont {Mattelaer}}, \bibinfo {author} {\bibfnamefont {H.~S.}\ \bibnamefont {Shao}}, \bibinfo {author} {\bibfnamefont {T.}~\bibnamefont {Stelzer}}, \bibinfo {author} {\bibfnamefont {P.}~\bibnamefont {Torrielli}},\ and\ \bibinfo {author} {\bibfnamefont {M.}~\bibnamefont {Zaro}},\ }\bibfield  {title} {\bibinfo {title} {{The automated computation of tree-level and next-to-leading order differential cross sections, and their matching to parton shower simulations}},\ }\href {https://doi.org/10.1007/JHEP07(2014)079} {\bibfield  {journal} {\bibinfo  {journal} {JHEP}\ }\textbf {\bibinfo {volume} {07}},\ \bibinfo {pages} {079}},\ \Eprint
  {https://arxiv.org/abs/1405.0301} {arXiv:1405.0301 [hep-ph]} \BibitemShut {NoStop}%
\bibitem [{\citenamefont {Stelzer}\ and\ \citenamefont {Long}(1994)}]{Stelzer:1994ta}%
  \BibitemOpen
  \bibfield  {author} {\bibinfo {author} {\bibfnamefont {T.}~\bibnamefont {Stelzer}}\ and\ \bibinfo {author} {\bibfnamefont {W.~F.}\ \bibnamefont {Long}},\ }\bibfield  {title} {\bibinfo {title} {{Automatic generation of tree level helicity amplitudes}},\ }\href {https://doi.org/10.1016/0010-4655(94)90084-1} {\bibfield  {journal} {\bibinfo  {journal} {Comput. Phys. Commun.}\ }\textbf {\bibinfo {volume} {81}},\ \bibinfo {pages} {357} (\bibinfo {year} {1994})},\ \Eprint {https://arxiv.org/abs/hep-ph/9401258} {arXiv:hep-ph/9401258} \BibitemShut {NoStop}%
\end{thebibliography}%

\end{document}